\documentclass[twocolumn]{aastex631}
\usepackage{ascii}
\usepackage{amsmath}
\usepackage{multirow}
\usepackage{graphicx}

\renewcommand{\thefootnote}{\alph{footnote}}

\newcommand{\desk}{{\small DESK}}
\newcommand{\dusty}{{\small DUSTY}}
\newcommand{\moreofdusty}{{\small MOD}}
\newcommand{\spitz}{Spitzer}
\newcommand{\kms}{~km\,s$^{-1}$}
\newcommand{\mum}{~$\mu$m}

\newcommand{\vgas}{\ensuremath{v_{\rm gas}}}
\newcommand{\vdust}{\ensuremath{v_{\rm dust}}}
\newcommand{\vdrift}{\ensuremath{v_{\text{drift}}}}
\newcommand{\rhod}{\ensuremath{\rho_{\text{d}}}}
\newcommand{\rhog}{\ensuremath{\rho_{\text{g}}}}

\newcommand{\FD}{\ensuremath{\mathfrak{F}_{\text{D}}}}
\newcommand{\mlr}{\dot{M}_{\rm gd}}
\newcommand{\rgd}{r_{\rm gd}}

\graphicspath{{./}{}}

\received{October 4, 2024}
\revised{December 19, 2024}
\accepted{January 2, 2025}

\submitjournal{ApJ}
\shorttitle{equatorial enhancement in the dustiest}
\shortauthors{Goldman et al.}

\begin{document}

\title{Equatorial Enhancement in the Dustiest OH/IR Stars in the Galactic Bulge}

\correspondingauthor{Steven Goldman}
\email{sgoldman@stsci.edu}

\author[0000-0002-8937-3844]{Steven R. Goldman}
\affil{Space Telescope Science Institute, 3700 San Martin Drive, Baltimore, MD 21218, USA}

\author[0000-0002-1272-3017]{Jacco Th. van Loon}
\affiliation{Lennard-Jones Laboratories, Keele University, ST5 5BG, UK}

\author[0000-0003-4870-5547]{Olivia C. Jones}
\affiliation{UK Astronomy Technology Centre, Royal Observatory, Blackford Hill, Edinburgh EH9 3HJ, UK}

\author[0000-0002-5797-2439]{Joris A. D. L. Blommaert}
\affiliation{Astronomy and Astrophysics Research Group, Department of Physics and Astrophysics, Vrije Universiteit Brussel, Pleinlaan 2, B-1050 Brussels, Belgium}

\author[0000-0003-2723-6075]{Martin A. T. Groenewegen}
\affiliation{Koninklijke Sterrenwacht van Belgi\"e, Ringlaan 3, B--1180 Brussels, Belgium}

\begin{abstract}
We have detected the 10\mum\ silicate feature and the 11.3\mum\ crystalline forsterite feature in absorption in 21 oxygen-rich AGB stars in the Galactic bulge. The depths of the 10\mum\ feature indicate highly obscured circumstellar environments. The additional crystalline features may suggest either an extended envelope or dust formation in a high-density environment. We have also modeled the spectral energy distributions of the sample using radiative transfer models, and compared the results to wind speeds measured using 1612 MHz circumstellar OH masers, as well as previous estimates of circumstellar properties. The 16 sources with measured pulsation periods appear on sequence D of the mid-IR period-luminosity relation, associated with the long secondary period. We suspect that all of these sources are in fact fundamental-mode pulsators. At least two sources appear on the fundamental mode sequence when accounting for the dust content. For the remainder, these sources are also likely fundamental-mode pulsators with extended envelopes. Taken as a whole, the high optical depths, crystalline features, discrepancies between observed and modeled wind speeds, pulsation periods longer than other fundamental-mode pulsators, and SED and pulsation properties similar to those with known equatorially enhanced circumstellar envelopes (e.g. OH\,26.5+0.6 and OH\,30.1--0.7) lead us to believe that these sources are likely to be equatorially enhanced.
\end{abstract}

\section{Introduction}

Thermally pulsing asymptotic giant branch (TP-AGB) stars contribute a tremendous amount of material back to the interstellar medium (ISM). They can lose up to 10$^{-4}$ $M_{\odot}$ yr$^{-1}$ and may collectively be the dominant source of stellar dust \citep{Hoefner2018,Hoppe2022}. The onset and timescale of this phase is an important constraint for stellar and chemical evolutionary models, and their dust injection rate is critical for tracing the origin of dust, particularly at high redshift.

AGB stars primarily have two types of surface chemistry, carbon- or oxygen-rich, which determine the type of dust produced. As AGB stars evolve, the surface chemistry is dictated by the initial and current mass (or stage of evolution on the AGB) and metallicity, which affect the efficiency and timescales of two important internal processes: third dredge-up \citep[TDU;][]{Herwig2005} and hot bottom burning \citep[HBB;][]{Boothroyd1993}. TDU events can occur above a certain limit in initial mass ($\sim$\,1.5\,$M_{\odot}$) and reach peak efficiency at around 2\,$M_{\odot}$ \citep{Ventura2018,Rees2024}. The convective intershell region reaches deep enough to bring carbon and $s$-process elements from the core to the surface, changing the surface chemistry from oxygen-rich to carbon-rich. For more massive stars ($M_{\rm init}$\,$\sim$\,4\,$M_{\odot}$), temperatures at the base of this convective envelope are high enough for hydrogen burning reactions to transform carbon into nitrogen through the CNO cycle, resulting in massive stars that retain a higher fraction of oxygen and an oxygen-rich atmosphere.

For AGB stars to produce dust, circumstellar material needs to reach sufficiently cool temperatures (T$\sim$1000\,K). As an AGB star evolves, gravity--opacity instabilities trigger large radial pulsations of the atmosphere \citep{Wood1999,Trabucchi2017}, allowing for the levitation of material out to large radii, where it cools and condenses into dust \citep{Gail1999,Karovicova2013}. At this stage, radiation pressure on the dust grains pushes them radially outward, and through momentum coupling the dust grains then drag along the surrounding (mostly hydrogen) gas \citep{Hoefner2008}. This mass loss has been shown to increase dramatically for pulsation periods from 300 to 500\,days after which it appears to plateau \citep{Vassiliadis1993}. At this point, the star reaches what is thought to be its most efficient stage of mass loss, a phase referred to as the ``superwind'' phase \citep{Iben1983}. These TP-AGB stars are sometimes referred to as extreme or x-AGB stars as a result of their higher dust content. They are also identified by their pulsation behavior, often referred to as Mira variables or long period variables \citep[LPVs;][]{Merrill1923,Wilson1942}. LPVs and Mira variables are often less optically obscured than x-AGB stars, and so these are not necessarily a subset of x-AGB stars. 

Most Mira variables pulsate in the fundamental mode on the $P$--$L$ sequence \citep{Wood1999,Trabucchi2017}. Some TP-AGB stars, however, have shown variability in another sequence referred to as the long secondary period (LSP, or sequence D). While the mechanism for this sequence has long remained unclear, it was shown that the LSP is consistently associated with unusual amounts of dust \citep{Wood2009}. One in three LPV stars is associated with this sequence \citep{Wood2000}. Recent evidence has shown that this phenomenon is likely linked to binarity \citep{Soszynski2021}. The proposed scenario is that a former planet orbiting an AGB star accretes enough material from the circumstellar envelope to become a brown dwarf or low-mass main-sequence star. This (sub)stellar companion is then observed to be enveloped in a dust cloud that, when hidden behind the AGB host, explains a secondary minimum that has been seen in mid-IR (but not optical) lightcurves; dusty companions of this type have been observed in evolved AGB stars like L$_2$ Pup \citep{Kervella2016} and $\pi$\ Gru \citep{Homan2020}.

\begin{figure*}
\includegraphics[width=\linewidth]{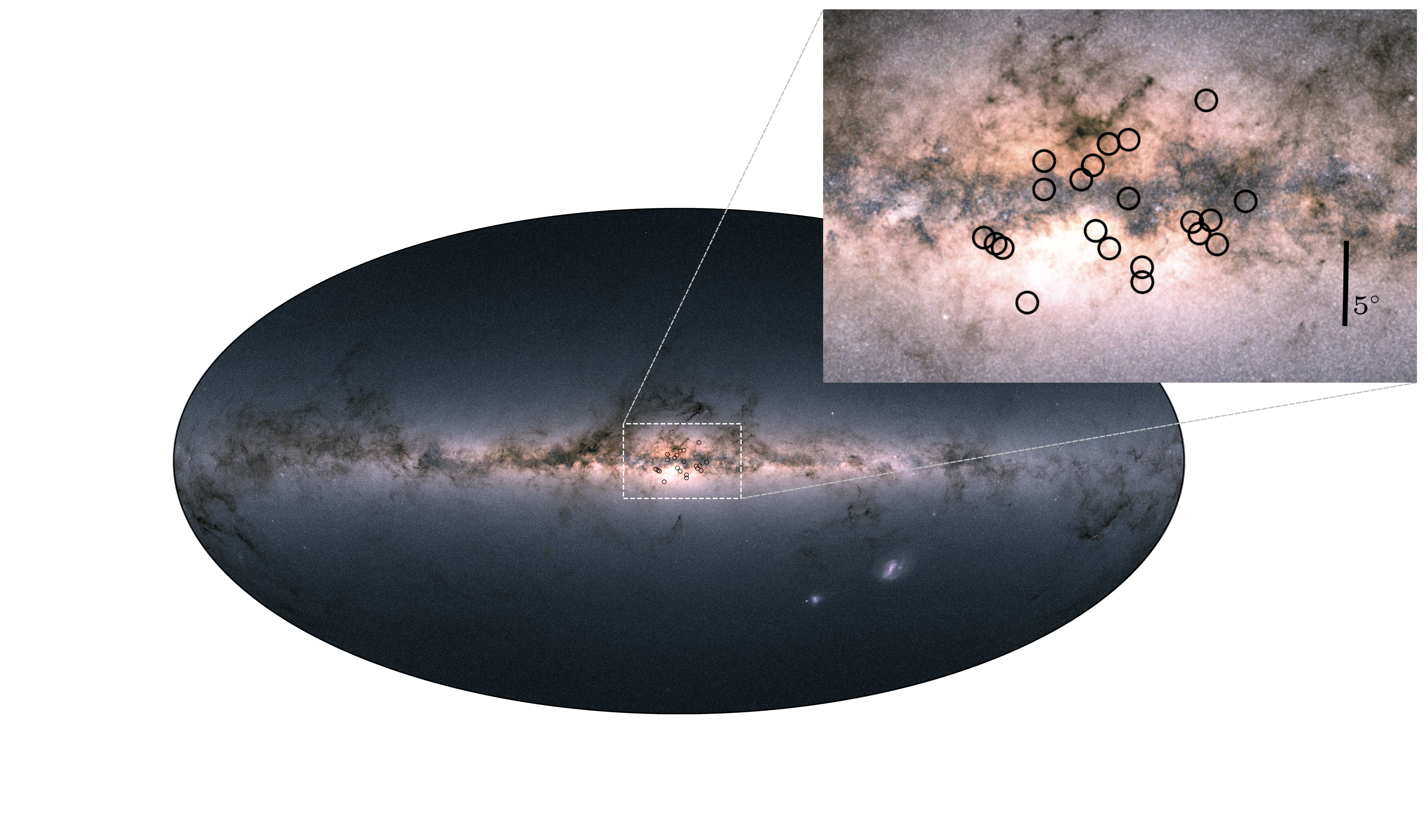} \vspace{-1.7cm}
\caption{The spatial distribution of the 21 GB AGB targets overlaid on a Hammer projection of the integrated sky map from Gaia Data Release 2 (Credit: ESA).\\}
\label{fig:gaia}
\end{figure*}

The dust surrounding evolved AGB stars makes them opaque at visible wavelengths but bright in the IR, as the radiation is absorbed and then reemitted by the dust. Within oxygen-rich AGB stars, the dust is expected to be made of silicates (olivines and pyroxenes) with some amounts of aluminum \citep{DellAgli2014,Jones2014} and iron, but a number of questions still surround the shape, structure, density, distribution of dust grain size, and whether iron is present in significant quantities \citep{McDonald2011c}. By modeling the overall shape of the spectral energy distributions (SEDs) of these stars, we can determine their bolometric luminosities as well as the level of dust obscuration. Dusty oxygen-rich AGB stars display a characteristic 10\mum\ silicate feature that is sensitive to changes in optical depth \citep{Trams1999,Sloan2008}. We can use this feature as a probe of circumstellar dust and measure the rate at which dust and mass are returned to the ISM.

Maser emission occurs around dusty oxygen-rich AGB stars and can be used to probe the kinematics of their outflows \citep{Sevenster2002}. The 1612\,MHz circumstellar OH maser occurs in the outer reaches of the circumstellar envelope, where outflowing material reaches its terminal expansion velocity. This emission, which is beamed radially, occurs on the near and far side of the star and typically shows two maser peaks. Given the large scale of the maser shell, we can assume that the star lies at the center of these peaks; half the peak separation provides a measure of the expansion velocity. These IR-bright AGB stars with detected OH maser expansion velocities are often referred to as OH/IR stars. The wind speeds provide another useful measurement of the efficiency of momentum transfer, measuring the result of the mass-loss mechanism, as opposed to estimating it from the mechanism's energy source (e.g.\ SED fitting).

\subsection{The Galactic Bulge}
It is now known that the Galactic bulge (GB) is a distinct Galactic component with different kinematics and composition from the Galactic plane and halo \citep[see review by][]{Minniti2007}. Understanding the environment and history of the bulge is critical to understanding its chemical enrichment, and the effect of that enrichment on AGB dust and outflows.

Observations have shown that the GB has a mass of 1.6\,$\times$\,$10^{10}$ $M_{\odot}$ \citep{Gerhard2006}, with both small \citep{Alard2001} and large bar structures \citep{Lopez-Corredoira2006}. There are also at least two main kinematic components within the GB that also have different metallicity distributions \citep[see ][and references therein]{Zoccali2018}. The metallicity peaks at nearly solar, with a sharp cutoff just above solar, and a tail toward lower metallicity \citep{Zoccali2003,Ness2016}.

Bulge stars have been found to be $\alpha$-enriched \citep{Nataf2011,Rosenfield2012}. This has been used to justify several bulge formation scenarios. With respect to the AGB population, $\alpha$-enrichment is expected to decrease the likelihood of carbon-rich chemistry in AGB stars by decreasing stellar lifetimes, resulting in fewer thermal pulses and less carbon dredge-up \citep{Karakas2014}. An absence of carbon-rich AGB stars in the bulge has also been confirmed observationally \citep{Sanders2023}.

\subsection{The OH/IR Population of the Galactic Bulge}
The GB provides an environment to test the dust production of a unique population of OH/IR stars. Previous works have shown that bulge OH/IR stars have a range of ages from 7 Gyr throughout the bulge as well as a population of younger OH/IR stars toward the inner bulge \citep{vanderVeen1990,vanLoon2003,Groenewegen2005,Blommaert2018}. These translate to initial masses covering the full age/mass range of AGB stars from 1 to 8\,$M_{\odot}$. Works have also shown distinct subsolar, solar, and supersolar populations \citep{Wood1983,Wood1998,GarciaPerez2018}. With a primarily metal-rich metallicity, however, we expect AGB chemical types that are dominated by oxygen-rich, as opposed to carbon-rich, chemistry. \citet{Marigo2020} isolated the initial masses (1.8--1.9\,$M_{\odot}$) of carbon stars in the solar neighborhood using a white dwarf initial--final mass relation. This is consistent with the narrow luminosity range of carbon-rich AGB stars in M31 \citep{Boyer2019,Goldman2022}. We expect oxygen-rich AGB stars in the GB to span initial masses outside of this narrow range. 

The 10\mum\ silicate feature, characteristic of oxygen-rich AGB stars, has been observed in the Galaxy in both emission and absorption. These discoveries were led primarily with the IRAS/Low-Resolution Spectrometer \citep{Olnon1986} and Spitzer/InfraRed Spectrograph \citep[IRS;][]{Sloan2003}. Within the Magellanic Clouds, the 10\mum\ silicate feature has been detected in emission and self-absorption, but not fully in absorption \citep{Trams1999,Dijkstra2005,Buchanan2006,Sloan2008,Woods2010,Jones2012,Ruffle2015a}. These represent different regimes of dust obscuration resulting from either absorption and reemission in an optically thin circumstellar environment ($\tau < 1$), or photon scattering in an optically thick environment ($\tau > 1$). In addition to circumstellar dust, interstellar dust may have an effect. Above the Galactic disk, however, the effects of interstellar attenuation are limited (Figure \ref{fig:gaia}), allowing us to accurately estimate the circumstellar extinction.

In this paper, we will investigate whether the GB evolved stars are as extreme as their previous photometric observations suggested, determine how these stars are being affected by their unique environment, and study their dust grain composition and pulsation behavior. Section 2 discusses the new and archival observations as well as our radiative transfer modeling method. Section 3 describes the results and comparative analysis with other comparable samples and results in the GB and Magellanic Clouds; section 4 is our concluding remarks. 

\section{Observations and Methods}

\subsection{Sample Selection}

Our GB sample is a subset of the sample targeted by \citet{Jimenez-Esteban2015}. The original sample was selected on the basis of having IR colors expected of bright oxygen-rich AGB stars ([12]--[25] $\gtrsim$ 0.75 mag) \citep{Jimenez-Esteban2006,Bedijn1987}. The sample was also chosen to be highly variable, selecting those with an IRAS variability index of VAR $>$ 50 \citep{Beichman1988}. From that sample, we selected 21 sources with measured expansion velocities from 1612 MHz OH maser observations. Two works recently have studied samples of Galactic OH/IR stars closer to the Galactic center \citep{Olofsson2022,Marini2023}; however, those samples share no overlap with our sample. Two of our sources were also studied in \citet{Blommaert2018}, which we will discuss further in \S\ref{sec:mlr}.

Pulsation periods have been measured for some of our sources \citep{vanderVeen1990,Jimenez-Esteban2015,Goldman2017,Groenewegen2022,Molnar2022}. Distances to individual sources are not as well known (c.f.\ Magellanic Clouds). Figure \ref{fig:gaia} shows the distribution of our AGB stars toward the Galactic center. The sample is generally unobscured by interstellar dust except for two sources, IRAS 17207$-$3632 and IRAS 17392$-$3020, which are spatially coincident with dust lanes. The photometry shown for reference in Figure \ref{fig:GB_seds} has been corrected for interstellar extinction; this is discussed further in \S\ref{sec:archival_data}.

\subsection{VLT {\rm N}-band Spectroscopy with VISIR}
We have used the Very Large Telescope (VLT)'s mid-IR imager and spectrograph \citep[VISIR;][]{Lagage2004}, mounted on the 8.2 m VLT Melipal telescope (UT3), to obtain long-slit $N$-band spectra of 21 sources within the GB. Our observations were taken on the night of 2017 June 18 using VISIR's low-resolution spectroscopy mode with a wavelength range of 8--13\mum\ and a resolving power ($\lambda$/$\Delta \lambda$) of 350 at 10\mum. The observations had a seeing in the optical that ranged between $0\rlap{.}\arcsec44$ and $1\rlap{.}\arcsec14$ and an average air mass of 1.4. A 10 minute integration was taken for each target. Sources within our sample that had lower signal-to-noise ratio (IRAS 17207$-$3632, IRAS 17351$-$3429, IRAS 17367$-$3633, IRAS 17392$-$3020, IRAS 17545$-$3056, IRAS 18092$-$2347, and IRAS 18195$-$2804) were observed a second time with the same length of integration. The observations were flux calibrated using the telluric standard star $\chi$ Sco taken the night before, and used chopping and nodding to remove the instrumental and atmospheric effects. The chopping used a frequency of 1.5 s$^{-1}$ and a chopping and nodding throw of 8\arcsec\ with an angle of zero degrees (east). This nodding strategy allowed for the targets to remain on the 32\arcsec\ long slit. For the first target, a slit width of 1\arcsec\ was used to test angular size within the slit, and had the same resolving power of 350. This was then switched to a slit width of $0\rlap{.}\arcsec75$, to decrease background noise. The observations were reduced using the VISIR pipeline (v4.3.1),\footnote{\href{https://www.eso.org/sci/software/pipelines/visir/visir-pipe-recipes.html} https://www.eso.org/sci/software/pipelines/visir/visir-pipe-recipes.html} extracted using EsoRex (v3.12.3), and are shown in Figure \ref{fig:GB_seds}. 

\begin{figure*}
  \begin{center}
     \includegraphics[height=3.9cm, trim={0.2cm 0.8cm 0.2cm 0}, clip]{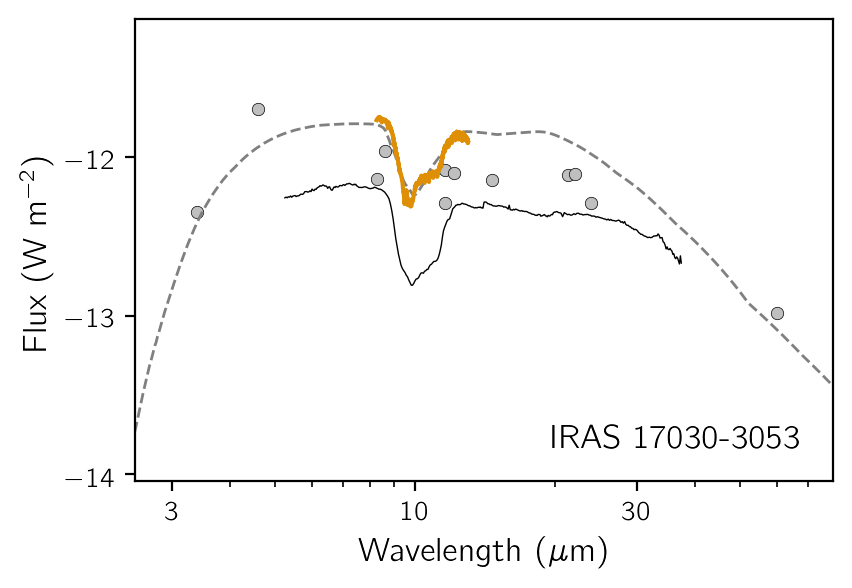}
     \includegraphics[height=3.9cm, trim={0.7cm 0.8cm 0.2cm 0}, clip]{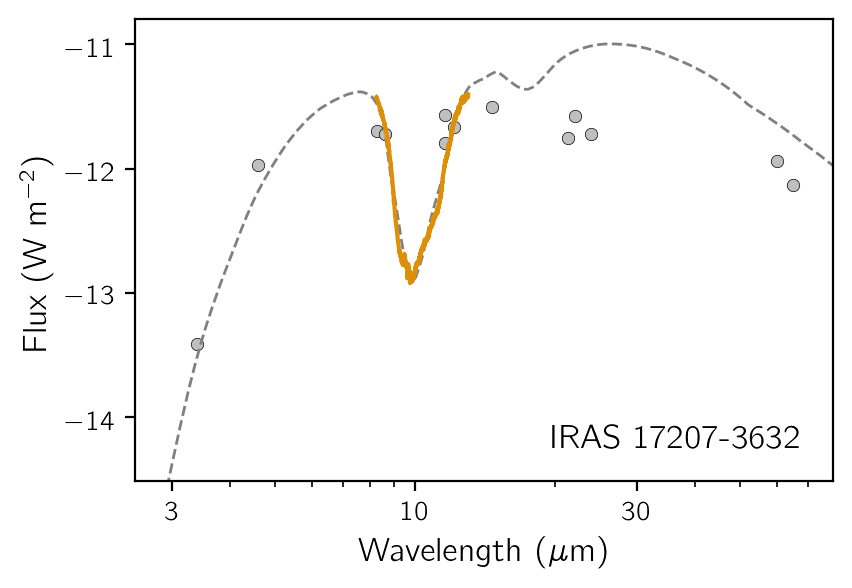}
     \includegraphics[height=3.9cm, trim={0.7cm 0.8cm 0.2cm 0}, clip]{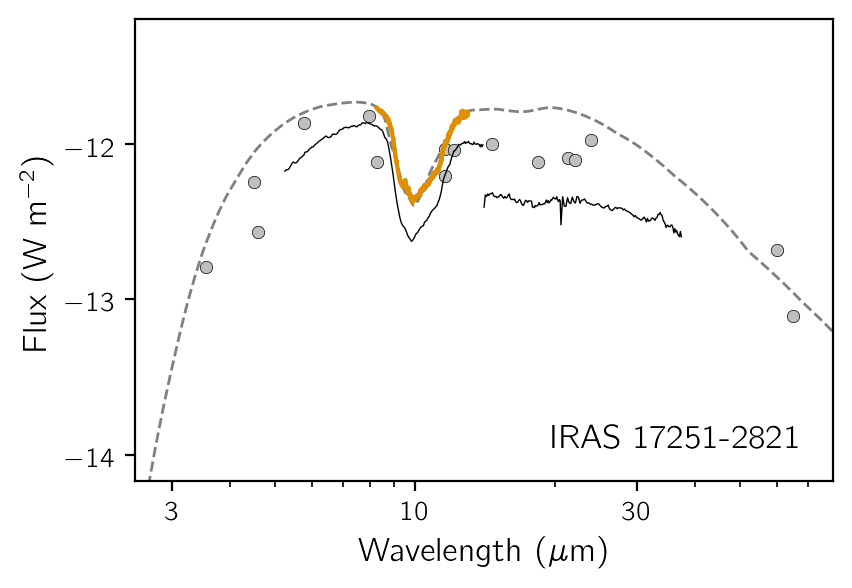}
     \includegraphics[height=3.9cm, trim={0.2cm 0.8cm 0.2cm 0}, clip]{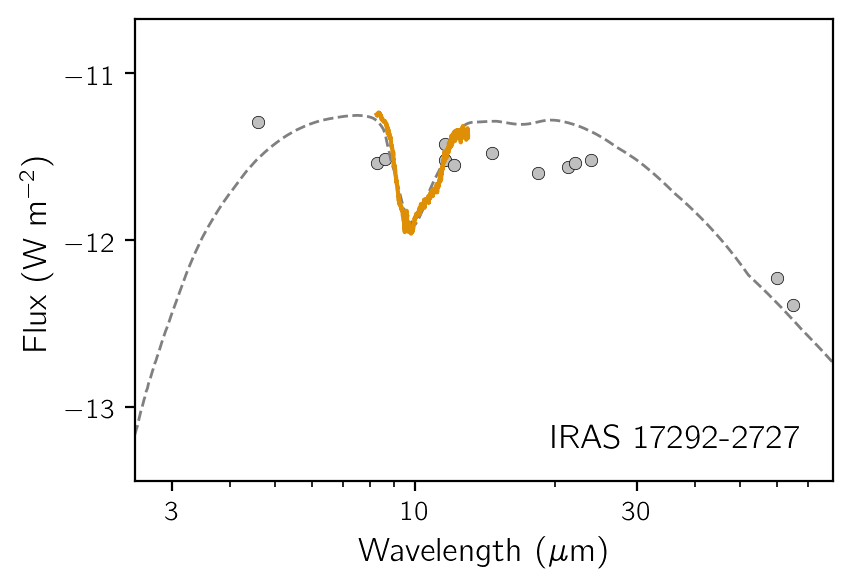}
     \includegraphics[height=3.9cm, trim={0.7cm 0.8cm 0.2cm 0}, clip]{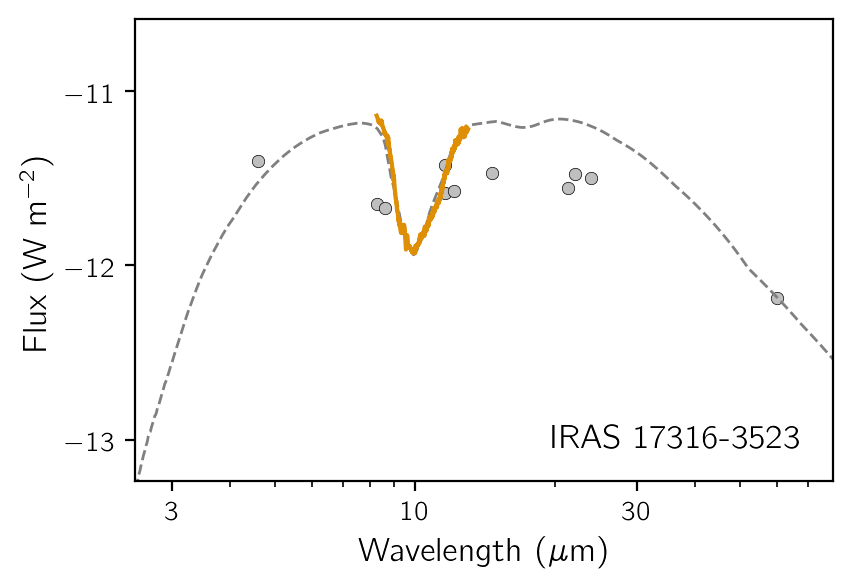}
     \includegraphics[height=3.9cm, trim={0.7cm 0.8cm 0.2cm 0}, clip]{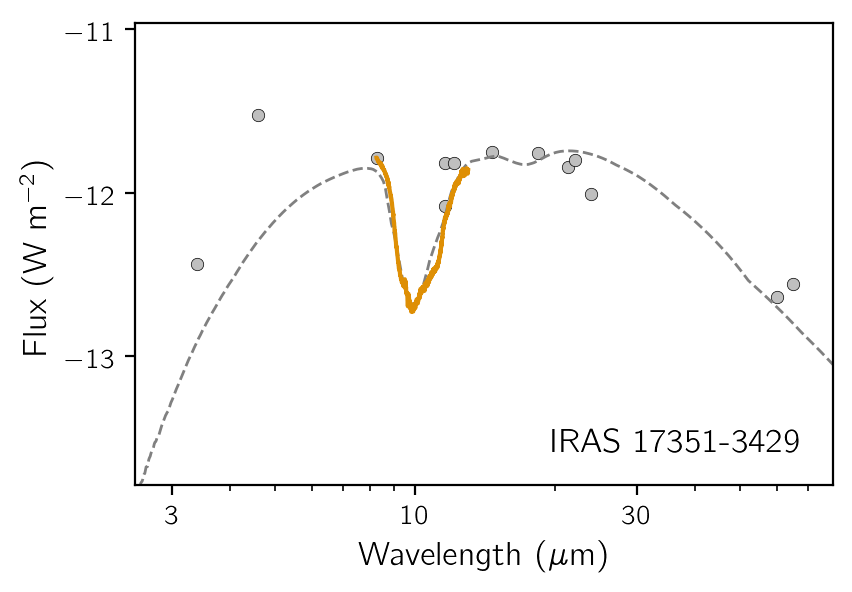}
     \includegraphics[height=3.9cm, trim={0.2cm 0.8cm 0.2cm 0}, clip]{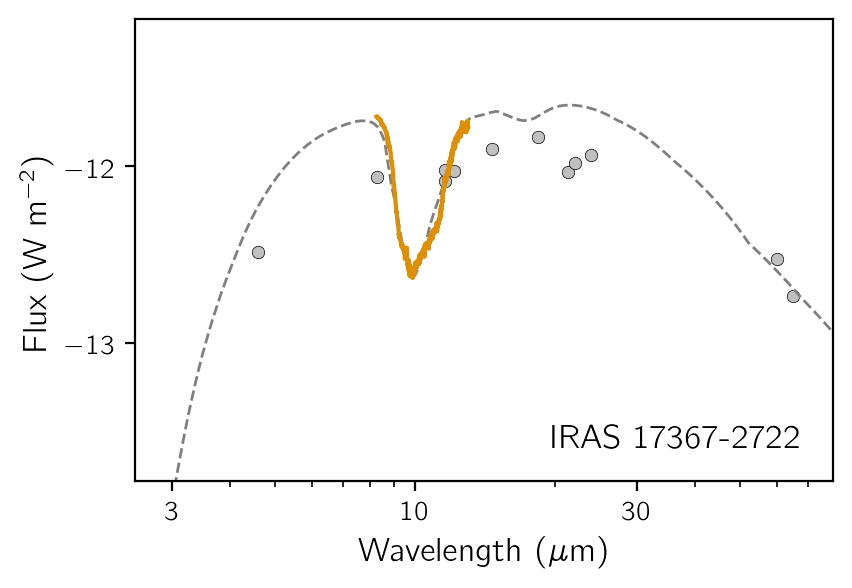}
     \includegraphics[height=3.9cm, trim={0.7cm 0.8cm 0.2cm 0}, clip]{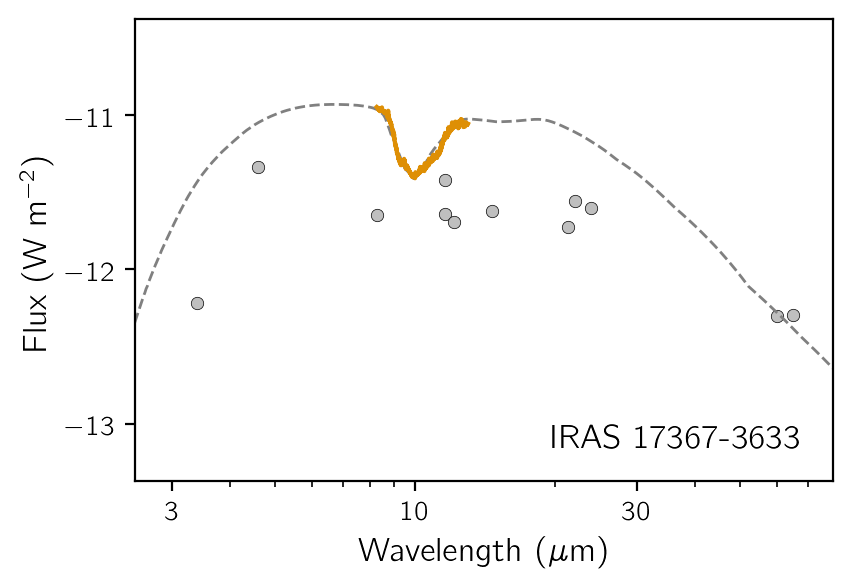}
     \includegraphics[height=3.9cm, trim={0.7cm 0.8cm 0.2cm 0}, clip]{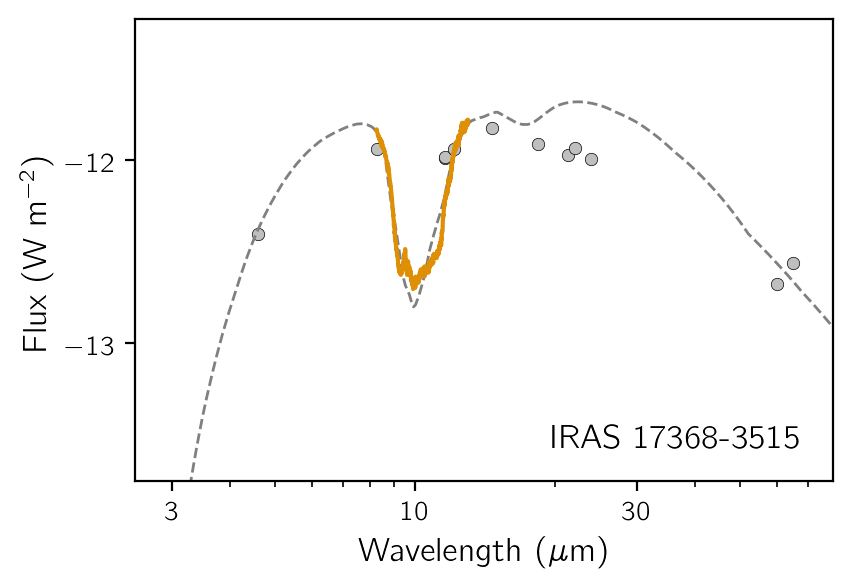}
     \includegraphics[height=3.9cm, trim={0.2cm 0.8cm 0.2cm 0}, clip]{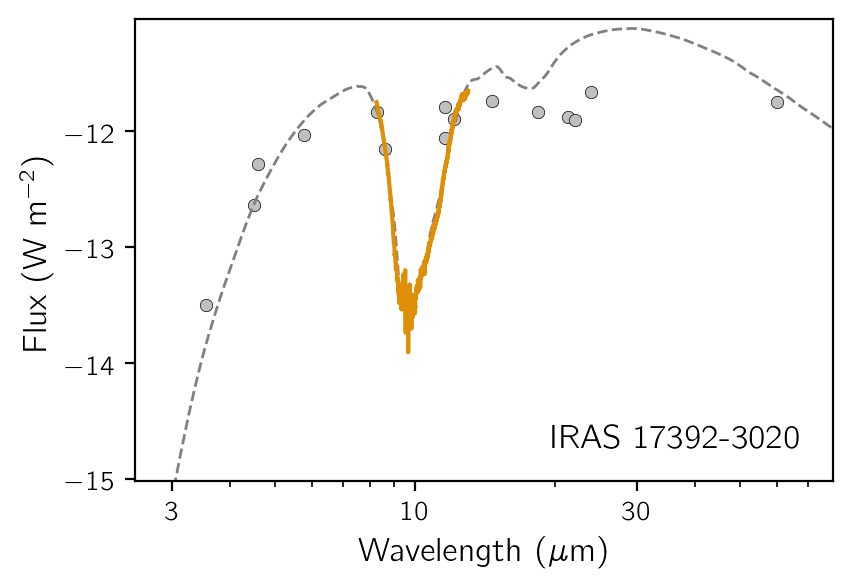}
     \includegraphics[height=3.9cm, trim={0.7cm 0.8cm 0.2cm 0}, clip]{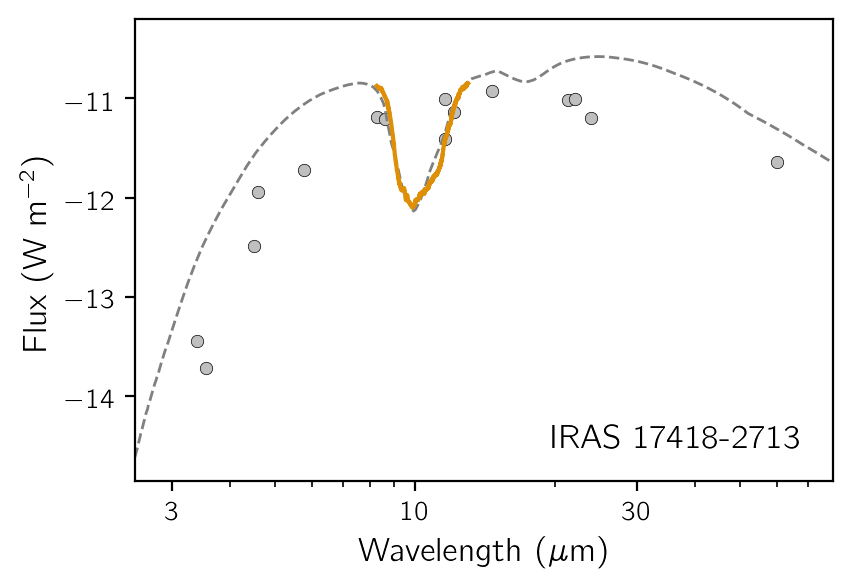}
     \includegraphics[height=3.9cm, trim={0.7cm 0.8cm 0.2cm 0}, clip]{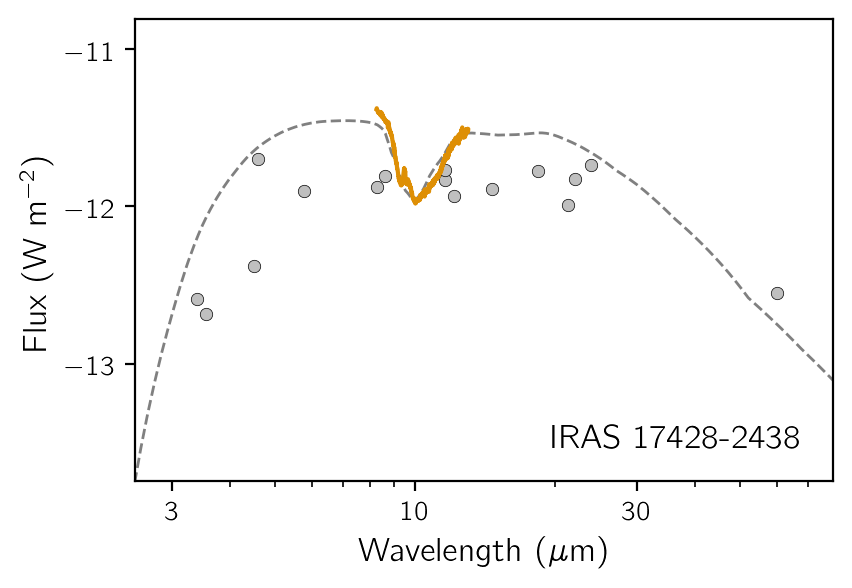}
     \includegraphics[height=4.375cm, trim={0.2cm 0.0cm 0.2cm 0}, clip]{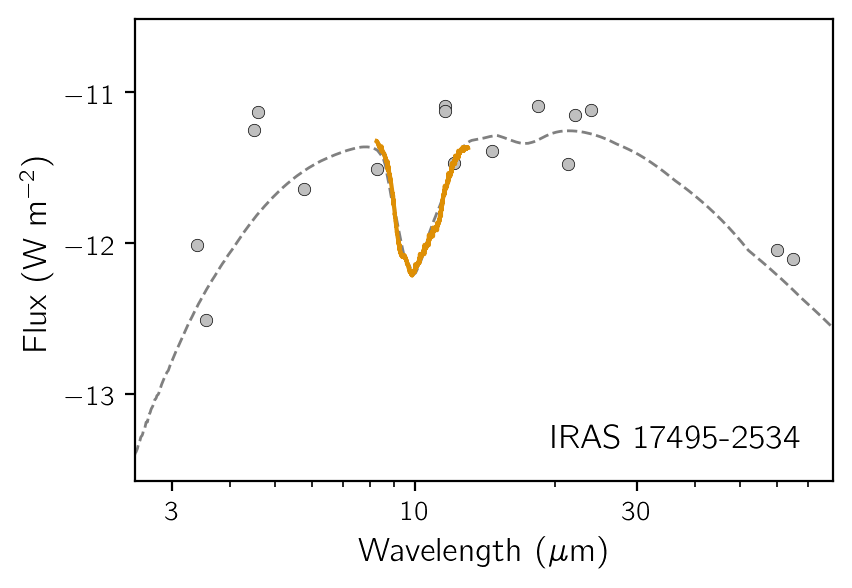}
     \includegraphics[height=4.375cm, trim={0.7cm 0.0cm 0.2cm 0}, clip]{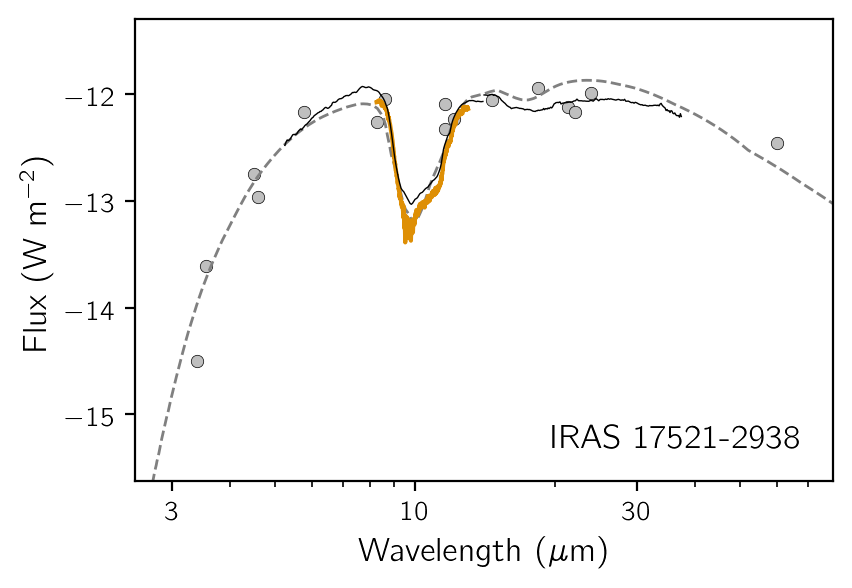}
     \includegraphics[height=4.375cm, trim={0.7cm 0.0cm 0.2cm 0}, clip]{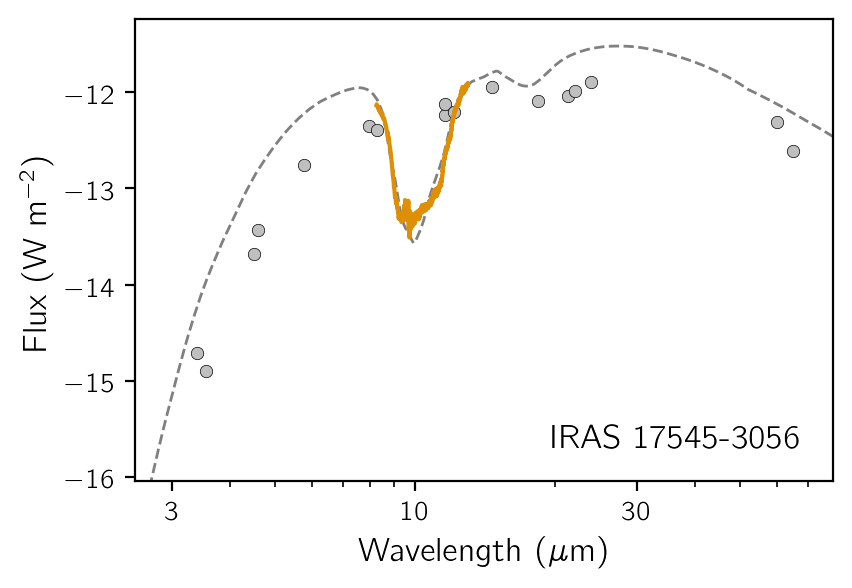}
   \end{center}
   \caption{The SED fitting of \dusty\ models (dashed line) to VISIR (orange) spectra for our sources within the GB. Also shown are the available \spitz /IRS spectra (in thin black) and available mid-IR photometry from \citet{Jimenez-Esteban2015}. Additional figures showing only the 10\mum\ region are shown in Figure \ref{fig:GB_seds_zoom_10}. The VISIR spectra shown in this figure are available as data behind the figure.}
   \label{fig:GB_seds}
\end{figure*}

\setcounter{figure}{1}
\begin{figure*}
  \begin{center}
     \includegraphics[height=3.9cm, trim={0.2cm 0.8cm 0.2cm 0}, clip]{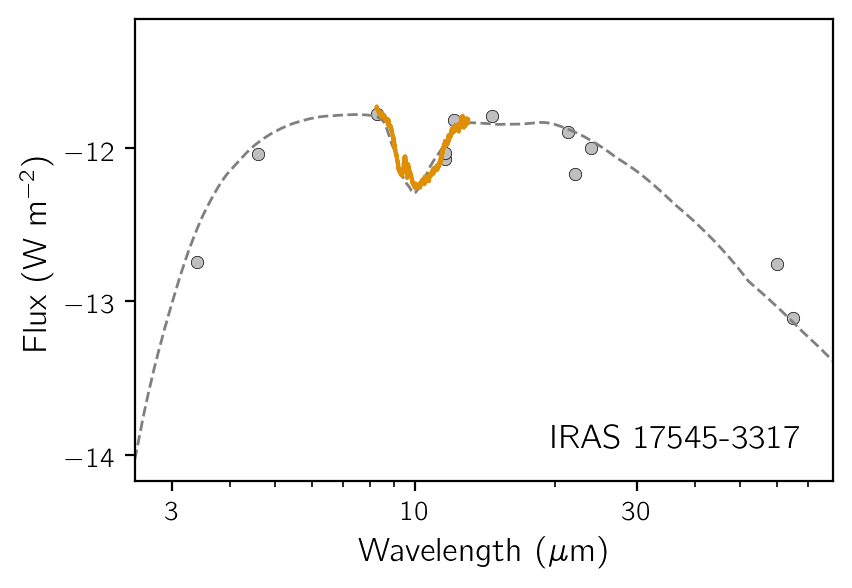}
     \includegraphics[height=3.9cm, trim={0.7cm 0.8cm 0.2cm 0}, clip]{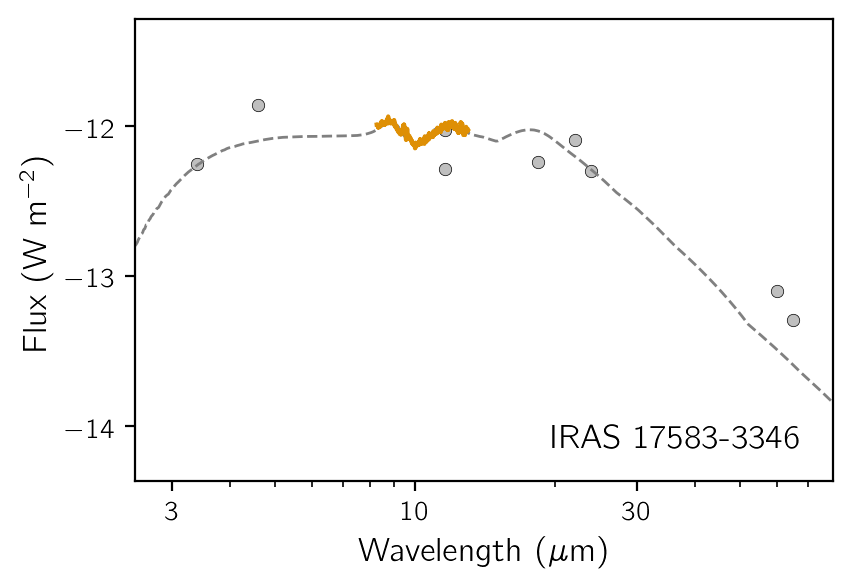}
     \includegraphics[height=3.9cm, trim={0.7cm 0.8cm 0.2cm 0}, clip]{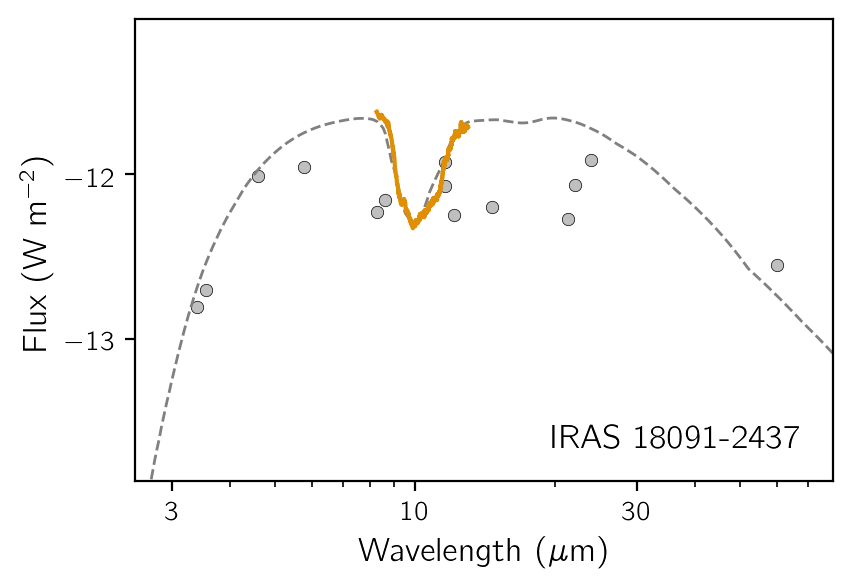}
     \includegraphics[height=4.3cm, trim={0.2cm 0.0cm 0.2cm 0}, clip]{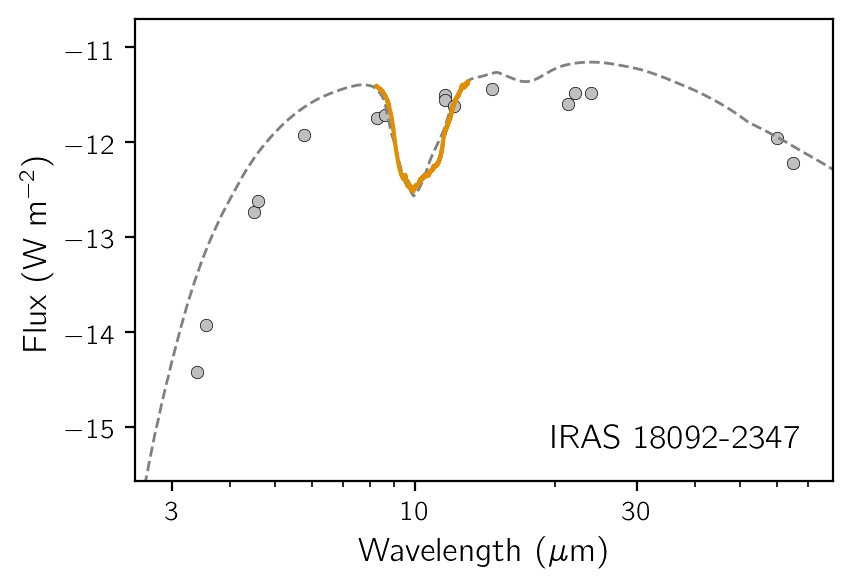}
     \includegraphics[height=4.3cm, trim={0.7cm 0.0cm 0.2cm 0}, clip]{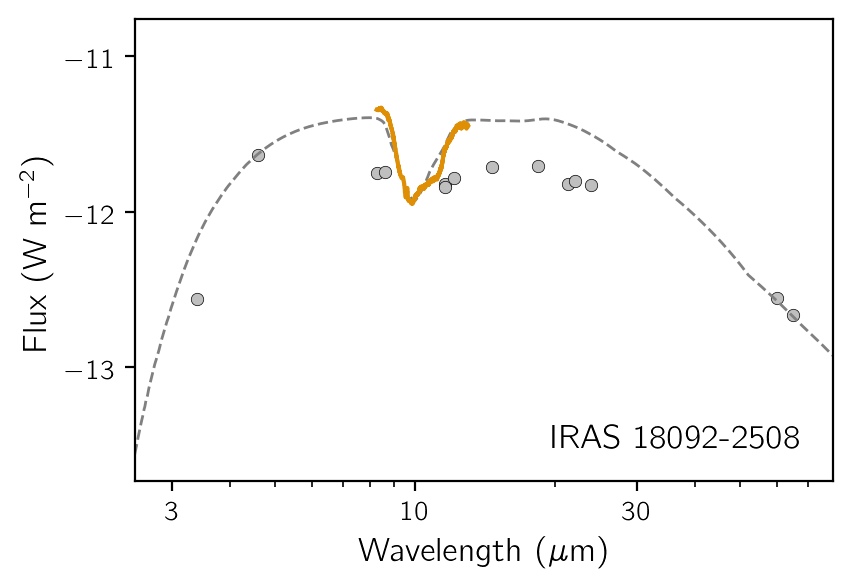}
     \includegraphics[height=4.3cm, trim={0.7cm 0.0cm 0.2cm 0}, clip]{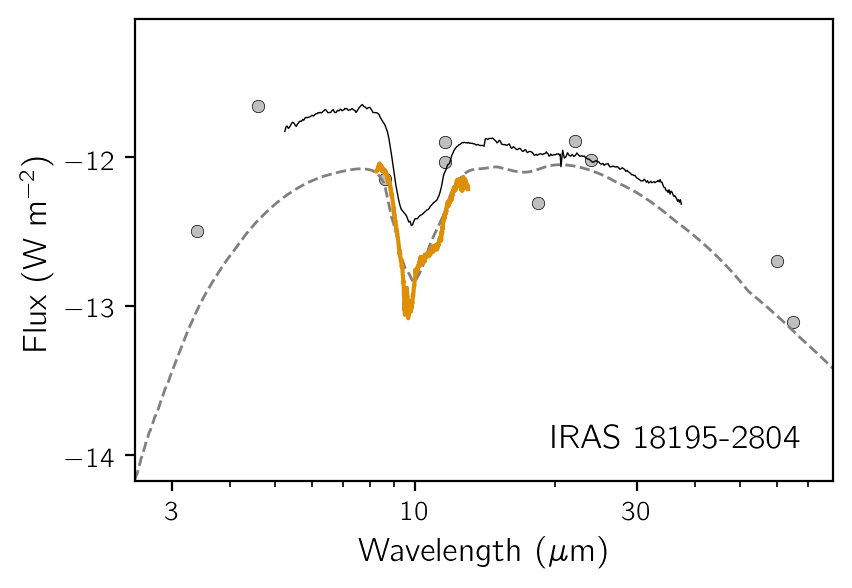}
   \end{center}
   \caption{continued}
\end{figure*}

\begin{figure*}
\includegraphics[height=6.0cm]{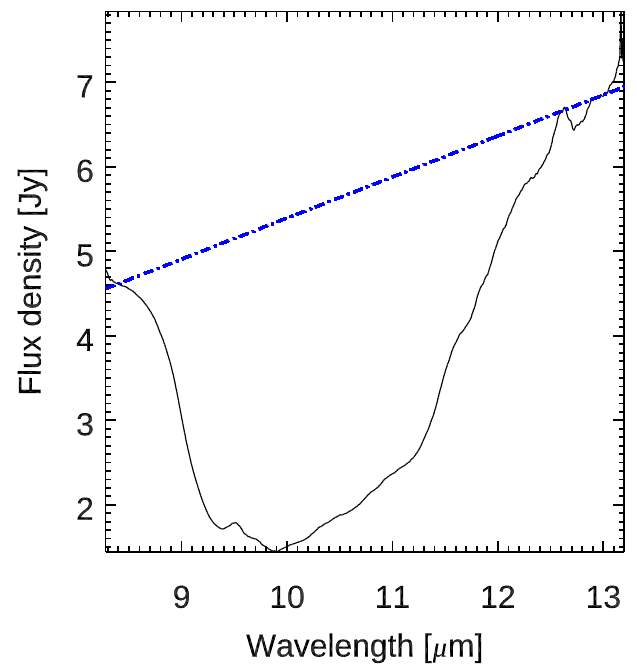}
\includegraphics[height=6.0cm]{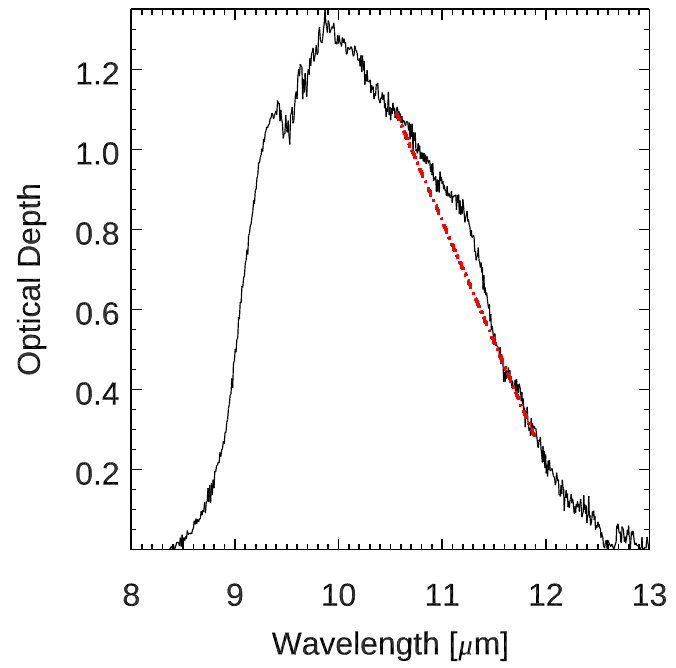}
\includegraphics[height=6.0cm]{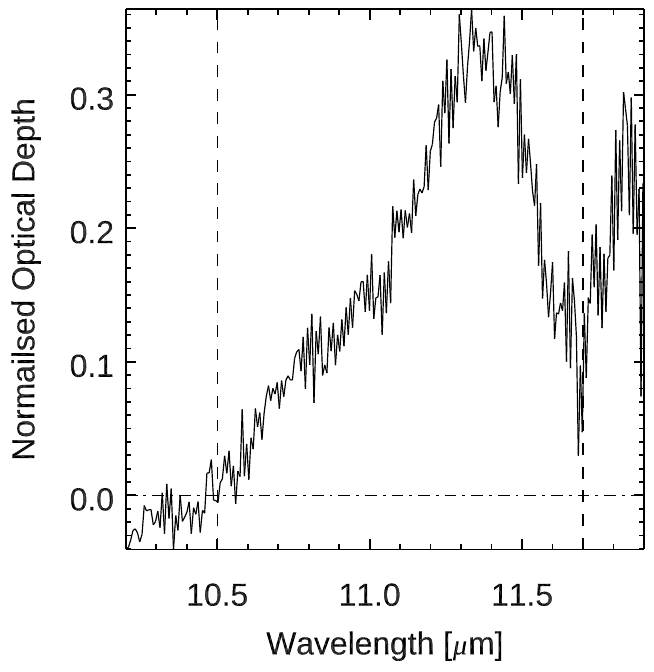}
\caption{The fitting of the 11.3\mum\ crystalline silicate feature of an example source (IRAS 17251$-$2821). Shown is the best-fit solution of the smoothed continuum (left), the optical depth (center), and the normalized optical depth (right), with the range over which the equivalent width ($W$\textsubscript{EQ}) was calculated shown with the dashed lines; $W$\textsubscript{EQ} is shown in Table \ref{table:sed_fitting_results}.}
\label{fig:crystalline_source}
\end{figure*}

\newcommand{\blommaert}{(a)}
\newcommand{\goldman}{\rlap{$^\dagger$}}
\newcommand{\groenewegen}{(b)}
\newcommand{\molnar}{(c)}
\newcommand{\vanderveen}{(d)}

\begin{deluxetable*}{lccrcrr}
\tablewidth{\columnwidth}
\tabletypesize{\small}
\tablecolumns{7}
\tablecaption{The GB sample properties and results of fitting the SEDs with radiative transfer models. \\ \label{table:sources}}
\tablehead{
\colhead{Target} &
\colhead{R.A.} &
\colhead{Decl.} &
\colhead{$A$\textsubscript{v}} &
\colhead{$P$} &
\colhead{W\textsubscript{EQ}} &
\colhead{$v$\textsubscript{exp\,OH}} \\
 &
 &
 &
{\footnotesize (mag)} &
{\footnotesize (d)} &
{\footnotesize ($\mu$m}) &
{\footnotesize (km\,s$^{-1})$}
}
\startdata
IRAS 17030$-$3053 & 17:06:14.061 & $-$30:57:38.284 &      & 780 \vanderveen & 0.26 & 15.3 \\
IRAS 17207$-$3632 & 17:24:07.254 & $-$36:35:40.444 & 19.6 &  926 \groenewegen & 0.09 & 16.1 \\
IRAS 17251$-$2821 & 17:28:18.601 & $-$28:24:00.332 & 4.1  &  698 \molnar \goldman & 0.10 & 16.0 \\
IRAS 17292$-$2727 & 17:32:23.559 & $-$27:30:01.114 & 5.6  &  919 \molnar \goldman & 0.13 & 17.6 \\
IRAS 17316$-$3523 & 17:34:57.489 & $-$35:25:52.508 & 7.9  &  646 \molnar \goldman & 0.05 & 12.3 \\
IRAS 17351$-$3429 & 17:38:26.269 & $-$34:30:40.721 & 5.3  &  \llap{1}062 \molnar \goldman & 0.13 & 20.0 \\
IRAS 17367$-$2722 & 17:39:52.423 & $-$27:23:31.824 & 5.0  &  & 0.12 & 7.7  \\
IRAS 17367$-$3633 & 17:40:07.616 & $-$36:34:41.256 & 4.0  & 777 \molnar \goldman & 0.09 & 14.5 \\
IRAS 17368$-$3515 & 17:40:12.970 & $-$35:16:40.728 & 3.6  &  & 0.16 & 13.5 \\
IRAS 17392$-$3020 & 17:42:30.536 & $-$30:22:07.368 & 17.0 &  & 0.05 & 21.1  \\
IRAS 17418$-$2713 & 17:44:58.730 & $-$27:14:42.830 & 7.0  &  & 0.10 & 15.2  \\
IRAS 17428$-$2438 & 17:45:56.939 & $-$24:39:57.841 & 5.7  & 695 \molnar & 0.05 & 13.8 \\
IRAS 17495$-$2534 & 17:52:39.534 & $-$25:34:39.174 & 10.5 & \llap{1}110 \molnar & 0.09 & 16.0 \\
IRAS 17521$-$2938 & 17:55:21.795 & $-$29:39:12.920 & 2.2  & 562 \blommaert & 0.14 & 16.8 \\
IRAS 17545$-$3056 & 17:57:48.413 & $-$30:56:25.721 & 2.9  & \llap{1}067 \groenewegen & 0.13 & 16.3  \\
IRAS 17545$-$3317 & 17:57:49.191 & $-$33:17:47.448 & 2.9  & 651 \molnar & 0.10 & 15.0  \\
IRAS 17583$-$3346 & 18:01:39.278 & $-$33:46:00.438 & 2.0  &  & $-$0.46 & 13.9 \\
IRAS 18091$-$2437 & 18:12:16.144 & $-$24:36:42.824 & 5.1  & 652 \molnar \goldman & 0.13 & 15.7 \\
IRAS 18092$-$2347 & 18:12:20.422 & $-$23:46:55.780 & 5.1  & \llap{1}430 \groenewegen & 0.14 & 17.4  \\
IRAS 18092$-$2508 & 18:12:21.866 & $-$25:07:20.571 & 4.4  & 630 \molnar & 0.17 & 14.5 \\
IRAS 18195$-$2804 & 18:22:40.214 & $-$28:03:08.643 & \llap{$\sim$}0.0 & 801 \groenewegen \goldman & 0.17 & 16.3 \\
\enddata
\vspace{0.2cm} $\dagger$ Comparable periods found by \citet{Goldman2017} using a similar dataset. 
\vspace{0.2cm} \newline \noindent  {\bf Note}---\,The interstellar extinction ($A_{\rm v}$) is calculated for each source using $E$($H$--$K$\textsubscript{s}) values within $4^{\prime}$ of each source from the \citet{Schultheis2014} 3D extinction map and the extinction law from \citet{Nishiyama2009} with an $A(K_{s})$ / $E(J-K) = 0.528$; IRAS 17030$-$3053 fell outside of the footprint of this map and is left blank. These values show no noticeable correlation with the measured optical depth. Pulsation periods ($P$) are shown where available. Equivalent widths for the 11.3\mum\ crystalline silicate feature (W\textsubscript{EQ}) are shown as well as the OH maser wind speeds ($v$\textsubscript{exp\,OH}). Positions are from WISE \citep{Cutri2012}. \citet{Whitelock1991} found the \citet{vanderVeen1990} periods to be too long in several cases due to too few epochs.
\tablerefs{\blommaert\ \citet{Blommaert2018}, \groenewegen\ \citet{Groenewegen2022}, \molnar\ \citet{Molnar2022}, \vanderveen\ \citet{vanderVeen1990}.}
\end{deluxetable*}

\subsection{Archival Data}
\label{sec:archival_data}
For four of our target sources we have archival spectra covering a range of 5.2--38\mum\ from the IRS \citep{Lebouteiller2011} on board the \spitz\ Space Telescope \citep{Werner2004}; these are shown in Figure \ref{fig:GB_seds}. The data were taken in either late March of 2004 or 2005. The silicate profiles appear unchanged when compared to our new data. IRAS 17251$-$2821 shows disjointed spectra between Short-Low (5.2--14\mum, slit width $3\rlap{.}\arcsec6$) and Long-Low (14.0--38\mum, slit width 10\rlap{.}\arcsec5) which is likely from a data reduction sensitivity limit in wavelength toward the edges of the passband. 

In addition to spectra, the sample has been observed in a variety of surveys. Within Figure \ref{fig:GB_seds} we have included (for reference) available IR photometry compiled by \citet{Jimenez-Esteban2015}. The photometry are corrected for interstellar extinction using the extinction law from \citet{Fitzpatrick1999} and the $A_{V}$ listed in Table \ref{table:sources}. As these stars are bright in the IR, some of the photometry is above the saturation limits of the detectors (particularly for the Wide-field
Infrared Survey Explorer, WISE) and may not be reliable.

In addition to IR data, our sample has also been observed in the submillimeter/radio. The sample was selected because, in addition to the sources' variability and reddened color, each source has a measured expansion velocity from 1612 MHz OH maser emission. These were detected with various instruments \citep{teLintel1991,David1993,Sevenster1997} and compiled by \citet{Jimenez-Esteban2015}. This data, in concert with the mass-loss rates that we have estimated, gives us a better understanding of the kinematics of the stellar outflows. Two sources, IRAS 17251$-$2821 and IRAS 17521$-$2938, have also been detected in CO using the Atacama Pathfinder EXperiment telescope \citep{Blommaert2018}; these include the CO (3--2) and (2--1) transitions. Using this data, the authors determined expansion velocities and mass-loss rates by modeling the shape of these lines, which were comparable to previous results from OH maser emission and mass-loss rates derived from modeling the SED (presented later in Table \ref{table:CO_comparison}).

\subsection{SED Fitting of the VISIR Spectra}
We fitted our VISIR spectra from 8.26 to 13\mum\ using the \textsc{Dusty Evolved Star Kit} \citep[\desk;][]{Goldman2020}, which fits grids of radiative transfer models using a $\chi^2$ fitting technique. We use a grid of models\footnote{The model grids used (``galactic-bulge-OH-IR'' and ``galactic-bulge-OH-IR-mmrn'') available for download on Zenodo \citep[Version 8,][]{goldman_2024_14448621} or for fitting using the \desk; see more details at \href{https://dusty-evolved-star-kit.readthedocs.io/en/latest/grids.html}{https://dusty-evolved-star-kit.readthedocs.io/en/latest/grids.html}.} computed using the 1D radiative transfer code \dusty\ \citep{Elitzur2001}. The models compute the hydrodynamical solution for the density structure of the winds, taking into account the star's gravitational attraction. The calculated gas mass-loss rates scale in proportion to $L^{3/4}(r_{\textrm{gd}}\rho_{\textrm{s}})^{1/2}$, where the dust grain bulk density ($\rho_{\textrm{s}}$) is set as 3 g cm\textsuperscript{$-3$} and the gas-to-dust ratio ($r_{\rm gd}$) is 200. For the calculation of the luminosities we assume a distance to the GB of 8\,kpc. The models assume a standard MRN \citep{Mathis1977} grain size distribution; however, for several sources we found that increasing the maximum allowable grain size from 0.25\mum\ to 0.5\mum\ resulted in a better fit of the optical photometry. For these sources we use the modified-MRN grain size distribution; these sources are identified in Table \ref{table:sed_fitting_results}. Using the modified-MRN grain size distribution resulted in derived dust mass-loss rates within 15\% of our standard grid dust mass-loss rates. The model grid assumes a dusty envelope with 4\% metallic iron grains \citep{Ordal1988}. A nonzero value less than 10\% was found to be necessary in theoretical models of low-mass AGB stars \citep{Ventura2020}, and a 4\% fraction was found to be necessary for fitting another OH/IR star, OH 127.8+0.0 \citep{Kemper2002}. The remaining dust is varying fractions of oxygen-rich and crystalline silicates \citep{Ossenkopf1992,Jaeger1994} where the crystalline silicate fraction is set to 0\%, 1\%, 3\%, 6\%, 10\%, or 15\%, and the oxygen-rich silicate fraction is the remainder. Atmospheric COMARCS models \citep{Aringer2016} are used as the basis for the central star, with effective temperatures that range from 2600 to 3400 K in increments of 200 K. The models also have a range of inner dust temperatures from 600 to 1000 K in increments of 200 K and a set of 50 optical depths specified at 10\mum\ linearly spaced from 1 to 50. There is a known degeneracy between inner dust temperature and optical depth, but this does not significantly affect the derived mass-loss rates \citep{Beasor2016}. Additionally, changes in effective temperature result in a minimal change in mass-loss rate as opposed to changes in the optical depth. Another commonly used oxygen-rich grid of radiative transfer models is the GRAMS models \citep{Srinivasan2011, Sargent2011}, however these models do not reach high enough optical depths for our GB sources.

\begin{figure*}
\centering
\includegraphics[width=0.9\linewidth]{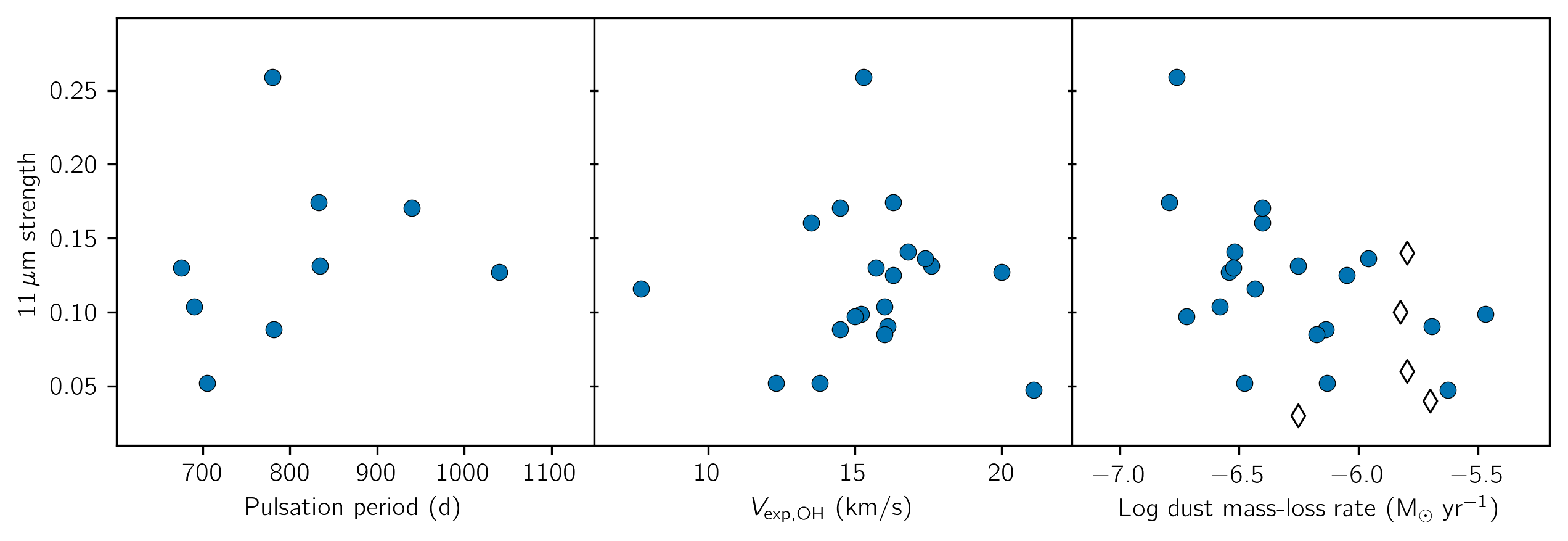}
\caption{The equivalent width of the 11.3\mum\ crystalline silicate feature with respect to other stellar measurements. The scatter in the equivalent width corresponds to a range in the crystalline fraction shown to reach above 10\% in some models \citep[see ][]{Vries2010}. The available data for the sample from \citet{Vries2010} are shown with diamonds. IRAS 17583$-$3346 is excluded from the figure as it falls well below the rest of the sample with a $W_{\rm EQ}=-0.46$.}
\label{fig:crystalline_fraction}
\end{figure*}

\begin{deluxetable*}{lrrrrrrrcr}
\tablewidth{\columnwidth}
\tablecolumns{11}
\tablecaption{The best-fit \small DUSTY model and derived parameters. This includes the effective temperature ($T$\textsubscript{eff}), inner dust temperature ($T$\textsubscript{inner}), predicted expansion velocities ($v$\textsubscript{exp}), optical depths specified at 10\mum\ ($\tau_{10}$), and the dust mass-loss rates (\textit{\.M$_{\rm dust}$}). Also shown is whether or not we used a modified-MRN (m-MRN) grain size distribution with a max grain size of 0.5\mum\ (instead of 0.25\mum). \\ \label{table:sed_fitting_results}}
\tablehead{
\colhead{Target} &
\colhead{$L$} &
\colhead{$T$\textsubscript{eff}} &
\colhead{$T$\textsubscript{inner}} &
\colhead{$v$\textsubscript{exp}} &
\colhead{$\tau_{10}$} &
\colhead{m-MRN} &
\colhead{\textit{\.M$_{\rm dust}$}}\\
&
{\footnotesize(10$^3$\,L\textsubscript{$\odot$})} &
{\footnotesize(K)} &
{\footnotesize(K)} &
{\footnotesize(km\,s\rlap{$^{-1})$}} &
{}&
&
{\footnotesize(10$^{-7}$}\,M\textsubscript{$\odot$}\,yr$^{-1}$)
}
\startdata
IRAS 17030$-$3053 & 5.7 & 3000 & 900 & 6.9 & 7.23 & Y & 1.7 \\
IRAS 17207$-$3632 & 26.4 & 3400 & 600 & 5.5 & 26.6 &  & 20.3 \\
IRAS 17251$-$2821 & 6.3 & 2600 & 900 & 6.6 & 11.3 & Y & 2.6 \\
IRAS 17292$-$2727 & 19.4 & 3200 & 800 & 7.8 & 11.3 &  & 5.6 \\
IRAS 17316$-$3523 & 23.8 & 3000 & 700 & 7.2 & 11.3 &  & 7.4 \\
IRAS 17351$-$3429 & 5.7 & 3400 & 600 & 4.3 & 10.3 &  & 2.9 \\
IRAS 17367$-$2722 & 7.0 & 3400 & 700 & 5.0 & 11.3 & Y & 3.7 \\
IRAS 17367$-$3633 & 39.7 & 3400 & 1000 & 12.0 & 10.3 &  & 7.3 \\
IRAS 17368$-$3515 & 6.3 & 3000 & 700 & 4.8 & 14.4 & Y & 4.0 \\
IRAS 17392$-$3020 & 19.4 & 3200 & 700 & 5.5 & 46.9 &  & 23.6 \\
IRAS 17418$-$2713 & 73.3 & 3200 & 600 & 7.4 & 20.5 &  & 34.1 \\
IRAS 17428$-$2438 & 11.6 & 3200 & 1000 & 8.9 & 9.27 & Y & 3.3 \\
IRAS 17495$-$2534 & 17.5 & 3400 & 600 & 5.7 & 10.3 &  & 6.7 \\
IRAS 17521$-$2938 & 3.8 & 3400 & 600 & 3.5 & 13.3 & Y & 3.0 \\
IRAS 17545$-$3056 & 7.7 & 3400 & 600 & 4.0 & 29.6 &  & 8.9 \\
IRAS 17545$-$3317 & 5.7 & 2600 & 900 & 6.8 & 8.25 & Y & 1.9 \\
IRAS 17583$-$3346 & 3.8 & 3400 & 800 & 6.2 & 3.16 & Y & 0.8 \\
IRAS 18091$-$2437 & 7.7 & 3400 & 800 & 6.1 & 9.27 & Y & 3.0 \\
IRAS 18092$-$2347 & 19.4 & 3400 & 600 & 5.3 & 14.4 & Y & 11.0 \\
IRAS 18092$-$2508 & 14.3 & 3400 & 800 & 7.5 & 7.23 & Y & 4.0 \\
IRAS 18195$-$2804 & 3.1 & 3200 & 700 & 4.3 & 11.3 &  & 1.6
\enddata
\end{deluxetable*}

\section{Results and Discussion}

\subsection{VISIR Spectra}
Figure \ref{fig:GB_seds} shows our observations spanning the 8--13\mum\ spectral range, intended to determine whether a sample of variable and highly reddened stars was as dusty as it appeared in the IRAS catalog. The VISIR spectra have indeed confirmed this. These observations have successfully detected and measured the 10\mum\ silicate feature in each source. This feature is highly sensitive to optical depth, and allows us to determine an accurate account of the line-of-sight dust content. This sample is particularly unique in that silicate absorption is seen in every source of our sample, which has yet to be detected in AGB stars outside of the Milky Way. The $N$-band spectral range also provides us with some information about the dust properties including the fraction of crystalline silicates.

The 11.3\mum\ crystalline silicate feature is a good tracer of the most recent mass loss of an AGB star. Evidence of crystalline silicates was previously discovered in the \spitz /IRS spectra of IRAS 17030$-$3053 and IRAS 17251$-$2821 \citep{Vanhollebeke2007,Chen2016}. We have now detected the 11.3\mum\ crystalline silicate feature in absorption on the shoulder of the 9.7\mum\ amorphous silicate feature in all but one of our sources. We have measured the crystalline fraction using the diagnostic tool developed by \citealt{Vries2010}. Here the optical depth of the 10\mum\ amorphous silicate feature is calculated using $F(\lambda)/F_{\rm cont}(\lambda) = {\rm e}^{-\tau}$. The 11.3\mum\ forsterite feature is then isolated by fitting a local continuum to the optical depth profile between 10.5--10.6\mum\ and 11.6--11.7\mum. Finally, the equivalent width of the forsterite feature is calculated from the normalized ($\tau(\lambda)/\tau_{\rm cont}(\lambda) -1$) profile by numerical integration from 10.5 to 11.7\mum. An example of the fitting method is shown in Figure \ref{fig:crystalline_source}. It is striking that we detected silicate absorption in all of our sources as well as the 11.3\mum\ silicate feature (except for one case). It remains unclear whether there is a correlation between the crystalline fraction and mass-loss rate \citep{Vries2010,Jones2012,Liu2017}. Figure \ref{fig:crystalline_fraction} shows the relationships of our crystalline fraction with expansion velocity, pulsation period, and mass-loss rate. We also include the sample from \citet{Vries2010} where data were available. These results cannot confirm or deny any clear relationships.

\subsection{SED Fitting Results}
\label{sec:sed_fitting}
We have fitted the VISIR spectra of each source to determine the properties of the samples' circumstellar dust and outflows. The results of the SED fitting are presented in Table \ref{table:sed_fitting_results} and show that these sources are extremely obscured with optical depths at 10\mum\ up to $\sim$\,47. The model fits were done on the VISIR spectra; however, as seen in Figure \ref{fig:GB_seds} the best-fit \dusty\ models trace the archival photometric data for these objects, albeit with some scatter. This is expected as these sources are variable and the measurements were taken at different epochs. Given the large uncertainties involved with SED fitting we forgo an in-depth discussion of the uncertainties in the modeling. We have, however, included a figure similar to Figure \ref{fig:GB_seds} showing the best-fit model and observations focused on the 10\mum\ region in the Appendix (Figure \ref{fig:GB_seds_zoom_10}).

Our SED fitting was done on a grid of models where the fraction of two different dust species, oxygen-rich silicates and crystalline silicates, was allowed to vary. The fraction of crystalline silicates was allowed to vary from 0 to 15\%. While we have detected the crystalline forsterite feature at 11.3\mum\ in our sample, this is not reflected as a significantly higher fraction of crystalline silicates in any of the best-fit models. Only two sources, IRAS 17392$-$3020 and IRAS 17545$-$3056, were found to have a nonzero crystalline fraction with a 6\% and 1\% fraction, respectively, for the best-fit model. These best-fit models that do include small fractions of crystalline silicates have very well fit silicate features (see the Appendix). The inclusion of these crystalline silicates deepens the feature at 11.3\mum, closer to what we see in our spectra, while also resulting in a deep narrow feature at $\sim$9.25\mum\ that is not observed in our spectra. This is however is near the region of atmospheric ozone that we are ignoring in our fitting. Additional testing of similar model grids with higher fractions of crystalline silicates (up to 50\%) did not result in improved fits (lower $\chi ^2$).

In several cases our models are not well fit to the available IR photometry from $\sim20$ to 30\mum . This includes photometry from the MSX6C Infrared Point Source Catalog \citep[E, 21.3\mum;][]{Egan2003}, the WISE catalog \citep[W4, 22.1\mum;][]{Wright2010}, and the IRAS Point Source Catalog \citep[25\mum;][]{Beichman1988}. For the WISE W4 band saturation issues begin at $\sim$12\,Jy, and at least seven of our sources are at or above this threshold. This may be limiting this photometry where it may otherwise have been a closer match to our best-fit models. The variability of these sources may also have contributed to a discrepancy between the data and the model. As the models in some of our sources overpredict the fluxes at these wavelengths, however, this is likely pointing to an inadequacy of the 1D modeling to reproduce geometric effects.

Previous works used the available photometry of the GB sample to model the overall SEDs \citep{Jimenez-Esteban2015,Goldman2017}. The available photometry was limited over the spectral range of the VISIR spectra, which is key to placing strong constraints on the optical depth of the source. In the previous works, the data were modeled and fit using a grid of 1D radiative transfer models from the \dusty\ code. Models were fit to median SEDs as the sources are variable and the observations were taken at different times. The large discrepancy with respect to those previous estimates of $\tau_{10}$ is not surprising as we now have a much better understanding of the optical depths of these sources. We have also calculated mass-loss rates that are higher than values previously found by both \citet{Jimenez-Esteban2015} and \citet{Goldman2017}; we will discuss possible explanations and uncertainties in the following sections.

\begin{table*}
\setlength{\tabcolsep}{0.15em}
\centering
\caption{A comparison of luminosities ($L$), expansion velocities ($v$\textsubscript{exp}), and dust mass-loss rates (\textit{\.M}\textsubscript{dust}) from SED-fitting and CO line modeling \citep{Blommaert2018}. Shown are results from SED fitting using the \textsc{Dusty Evolved Star Kit} \citep[\desk;][]{Goldman2020} and \textsc{More of Dusty} \citep[\moreofdusty;][]{Groenewegen2012} as well as the wind speeds from the maser observations \citep[][and references therein]{Jimenez-Esteban2015}, and CO line fitting by \citet{Blommaert2018}; also shown is the total mass-loss rate (\textit{\.M}\textsubscript{tot}). \label{table:CO_comparison}}
\begin{tabular}{lcccccccccccc}
\hline
\multirow{3.5}{*}{Target}& \hspace{3cm} & \multicolumn{2}{c}{\desk} & \hspace{3cm}& \multicolumn{2}{c}{\moreofdusty} & \hspace{3cm} & OH Maser & \hspace{3cm} & \multicolumn{3}{c}{CO Emission} \\

&& \multicolumn{2}{c}{$\overbrace{\rule{3.0cm}{0pt}}$}  &&
\multicolumn{2}{c}{$\overbrace{\rule{2cm}{0pt}}$}  &&
$\overbrace{\rule{0.75cm}{0pt}}$ &&
\multicolumn{3}{c}{$\overbrace{\rule{5cm}{0pt}}$}

\vspace{-0.2cm} \\
&& $L$ & \textit{\.M}\textsubscript{dust} && $L$ & \textit{\.M}\textsubscript{dust} && $v$\textsubscript{exp} && $v$\textsubscript{exp} & \textit{\.M}\textsubscript{tot}& \textit{\.M}\textsubscript{dust} \\

& & {\scriptsize($10^3$ $L$\textsubscript{$\odot$})} & {\scriptsize($10^{-7}$ M\textsubscript{$\odot$}\,yr$^{-1}$)} & & {\scriptsize($10^3$ $L$\textsubscript{$\odot$})} & {\scriptsize($10^{-7}$ M\textsubscript{$\odot$}\,yr$^{-1}$)} && {\scriptsize (km\,s$^{-1}$)} && {\scriptsize (km\,s$^{-1}$)} & {\scriptsize($10^{-5}$ M\textsubscript{$\odot$}\,yr$^{-1}$)} & {\scriptsize($10^{-7}$ M\textsubscript{$\odot$}\,yr$^{-1}$)}  \\
\hline\hline
IRAS 17251$-$2821  && 6.3 & 2.6  && 4.8 & 0.98 && 16.0 && 17.9 & 4.0 & 2.4 \\
IRAS 17521$-$2938 &&  3.8 & 3.0 && 4.1 & 2.0 && 16.8 && 18.2 & 7.0 & 4.6 \\

\hline
\end{tabular}
\end{table*}

\subsection{Luminosities and Initial Masses}
\label{sec:lum_mass}
We suspect that our sample is composed of a large range of initial masses. Previously, \citet{Jimenez-Esteban2015} split their broader sample into low- and high-luminosity groups, with the latter group representing AGB stars that are oxygen-rich as a result of HBB. Initial masses were estimated for larger samples that share some sources with our sample with ranges of 1--2.2\,$M_{\odot}$ \citep{vanderVeen1990} and 1.5--2\,$M_{\odot}$ \citep{Groenewegen2005}. \citet{Olofsson2022} estimated initial masses up to 4.3\,$M_{\odot}$ for their high-luminosity group of similar sources, but acknowledge that some of those may be foreground stars. Given our new data, we do not see a clear delineation between low- and high-luminosity sources in our sample, and thus we treat the sample as a single group. For sources for which we have pulsation periods, we have used PARSEC \citep{Bressan2012,Marigo2013} isochrones to visualize the range of initial masses $M_{\rm init}$=1--5\,$M_{\odot}$ (shown later in Figure \ref{fig:mbol_p_l_relation}). Our sample has luminosities that span beyond the full range of the end points of these isochrones, but with longer periods; this is discussed further in \S\ref{sec:pulsation_properties}. \\ 

\subsection{Expansion Velocities}
\label{sec:vexp}
Our average measured expansion velocity is more than double what is estimated with our best-fit model.\footnote{The \dusty\ code specifies that wind solutions apply only if the wind speeds exceed 5\kms, and as many of our sources are near or below this threshold, they should be used with caution.} On the large scales where OH maser emission occurs, outflows have been found to be primarily spherically symmetric \citep{Ramstedt2020} even as their morphology on smaller scales may vary significantly \citep[e.g.,][]{Decin2020}. As our modeled expansion velocities do not appear to represent the kinematics of the winds around the stars, this may suggest that dust densities are higher along our line of sight, or that they are exhibiting high drift velocities.\footnote{Similarly large deviations between modeled \dusty\ and OH maser expansion velocities are also seen in the OH/IR sample in the LMC \citep{Goldman2017}. The mass-loss rates estimated for that sample may similarly be overestimated as they show similar SED shapes and long pulsation periods.}

The differences in the modeled and observed wind speeds may stem from deviations from our assumed gas-to-dust ratio ($\rgd$) of 200 and our assumption of negligible drift velocity (\vdrift) between the gas and dust; this drift velocity can be expressed as $\vdrift = \vdust - \vgas$, where \vdust\ is the dust velocity, and \vgas\ is the gas velocity. The $\rgd$\ and $\vdrift$\ are exceedingly hard to isolate observationally and so we can analyze their impact together. Following the work by \citet{Sandin2020} we calculate the mass-loss ratio as
\begin{eqnarray}
\mlr=\frac{\rhog}{\rhod}\frac{\vgas}{\vdust}=\frac{\rgd}{\FD},\label{eq:masslossratio}
\end{eqnarray}
where \rhog\ is the gas density, \rhod\ is the dust density, and the gas-to-dust ratio is expressed as $\rgd=\rhog/\rhod$ and the drift factor as $\FD=\vdust/\vgas$. By rearranging the definition of the drift velocity, we can express the mass-loss ratio $\mlr$\ in terms of drift velocity and gas velocity in Eq. \ref{eq:masslossratiorearranged}:
\begin{eqnarray}
\mlr=\rgd*\frac{\vgas}{\vgas + \vdrift}\label{eq:masslossratiorearranged}
\end{eqnarray}

Under our assumption of negligible drift velocity $\mlr=\rgd$, where we assume $\rgd=200$.

Theoretical work has demonstrated that the drift velocity should be more significant in oxygen-rich AGB stars as opposed to carbon stars \citep{Mattsson2021}. Modeling by \citet{Sandin2023} confirms this, with average estimates of $\vdrift \sim 87$--310\kms\ and $\mlr \sim$ 33.8--57.8 for a solar-mass oxygen-rich AGB star with a luminosity range $L=7,000$--28,000\,$L_{\odot}$. Applying this $\mlr$ to our oxygen-rich sample would result in gas and total mass-loss rates $\sim 17$--29\% of our calculated values. Using a sample which included some of our targets, \citet{Blommaert2018} estimates a much slower drift velocity of 5\kms, and a gas-to-dust ratio of 100--400 leading to a $\mlr \sim 76$--305, assuming an average gas velocity $\vgas=16$\kms. This highlights the challenge in estimating mass-loss rates based purely on observational data. In either case, it is likely that we are overestimating mass-loss rates, and so they should be used with caution.

\subsection{Mass-loss Rates}
\label{sec:mlr}
Within our sample, we see high optical depths irrespective of luminosity. Our best-fit \dusty\ models allow us to estimate the dust mass-loss rate, which we then scale by a gas-to-dust ratio of 200 to calculate the gas mass-loss rates. Combining both the gas and dust mass-loss rates gives us our total mass-loss rates $\sim 10^{-4}$\,$M_{\odot}$\,yr$^{-1}$, characteristic of AGB stars in their superwind phase. These results assume spherical symmetry, which may not be the case for this sample, and so we expect this to represent more of an upper limit to the mass-loss rates. 

Two sources, IRAS 17251$-$2821 and IRAS 17521$-$2938, were detected in both 1612 MHz OH maser and CO line emission. We have compared our SED results with the SED and CO results of \citet{Blommaert2018} for these two sources as a benchmark (Table \ref{table:CO_comparison}). \citet{Blommaert2018} used a more sophisticated dynamical estimation of the outflows using the OH maser wind speeds and the CO emission features. We find slightly discrepant luminosities, but similar dust mass-loss rates as those derived from CO and SED analysis in \citet{Blommaert2018}.

\begin{figure}
\begin{center}
\includegraphics[width=\columnwidth]{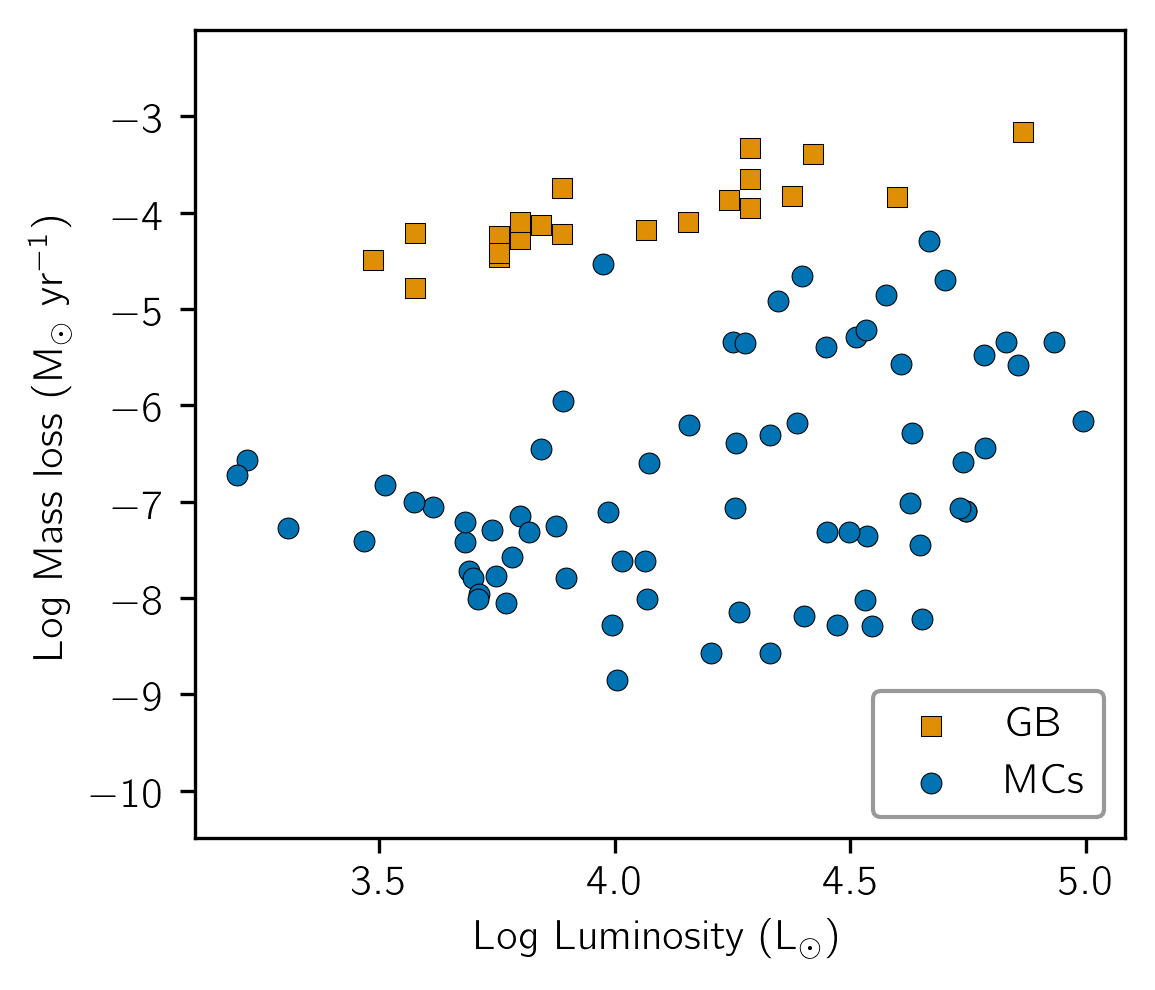}
\caption{Luminosities and total mass-loss rates of the GB sample and a comparison sample of dusty oxygen-rich AGB stars in the Magellanic Clouds from \citet{Groenewegen2018}. We suspect that our mass-loss rates may be overestimated as a result of equatorial enhancement.}
\label{fig:lum_and_mass_loss}
\end{center}
\end{figure}

The well-studied sample of dusty oxygen-rich AGB stars in the LMC gives us a less dusty comparison sample. Figure \ref{fig:lum_and_mass_loss} shows that, while our luminosities are similar to the dusty oxygen-rich AGB stars and red supergiants seen in the Magellanic Clouds, our total mass-loss rates are far higher. We expect our sample to have a metallicity distribution around solar, compared to approximately half-solar for the LMC comparison sample. The effects of metallicity on AGB outflows and evolution have been studied in nearby samples, but are not well constrained. Samples of AGB stars in nearby galaxies and globular clusters have shown that dust production \citep{Sloan2008,Sloan2010} and expansion velocity \citep{Marshall2004,Goldman2017} are limited in more metal-poor, oxygen-rich AGB stars. Additionally, the linear relationship between the ratio of oxygen- and carbon-rich AGB stars (C/M ratio) and metallicity shows a break around solar metallicity, but is still dominated by oxygen-rich AGB stars \citep{Boyer2019,Goldman2022}. It remains unclear how the total mass-loss rate is affected by changes in metallicity, but we suspect that the large difference in mass-loss rates are not dominated by metallicity effects. 

Fitting the IR and submillimeter data provides independent methods of determining stellar parameters that rely on different assumptions. \citet{Ramstedt2008} show that mass-loss rates determined by CO line emission tend to be higher than those measured by fitting SEDs with radiative transfer models, and state that it may be due to underestimates of the CO/H$_{2}$ ratio. As was mentioned above, mass-loss rates based on 1D radiative transfer models rely on assumptions of the gas-to-dust ratio, drift velocity, and spherical geometry. Taking into consideration these uncertainties, \citet{Blommaert2018} estimated an average uncertainty of 2 for their SED-derived dust mass-loss rates. As we will discuss in \S\ref{sec:equatorial_enhancement}, a nonspherical geometry may lead to an overestimated mass-loss rate by a factor of a few to $\sim100$ \citep{Decin2019}. 

Accompanying the uncertainties in the calculation of mass-loss rates are possible sample biases with respect to distance. None of our sources have reliable parallaxes from {\it Gaia} Data Release 3 \citep{Gaia2021}. This is not surprising as it has been shown that the convection cells in AGB stars result in a migration of the photocenters \citep{Chiavassa2018}. We expect that we may have an observational bias for closer stars as these stars were selected based on their IRAS colors and variability index, which may have missed more distant sources. All of the GB sources have maser peak flux densities on the order of hundreds of milliJansky and a range of radial velocities from $-180$ to 230\kms. However, there is one outlier, IRAS 17316$-$3523, with higher peak maser flux densities of 4.1 and 3.0 Jy for the blue and red maser peaks, respectively. This source may be closer in proximity, yet it does not have a significantly higher inferred luminosity than the rest of the sample.

\begin{figure*}
    \centering
    \includegraphics[width=0.49 \linewidth]{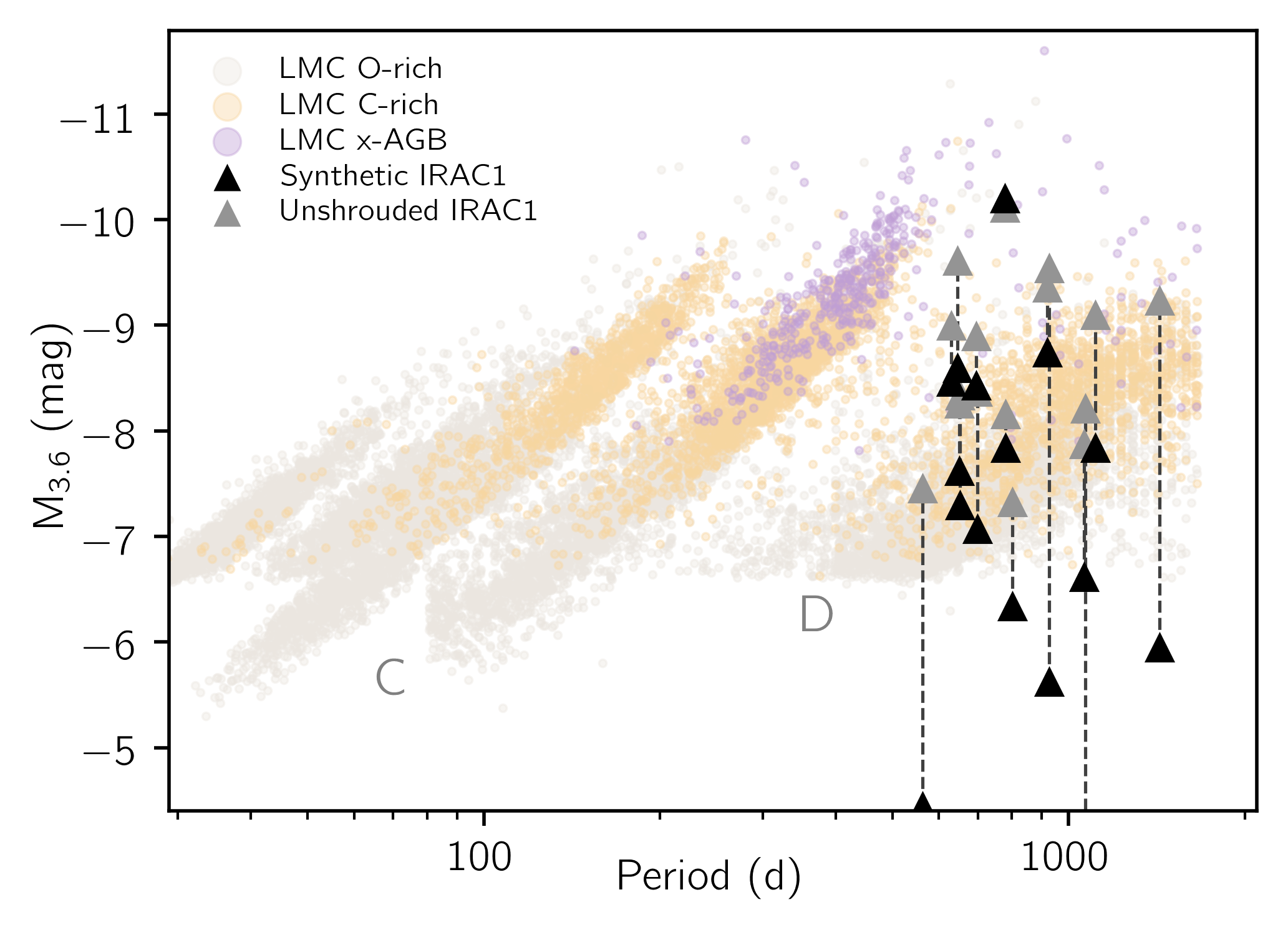}
    \includegraphics[width=0.49 \linewidth]{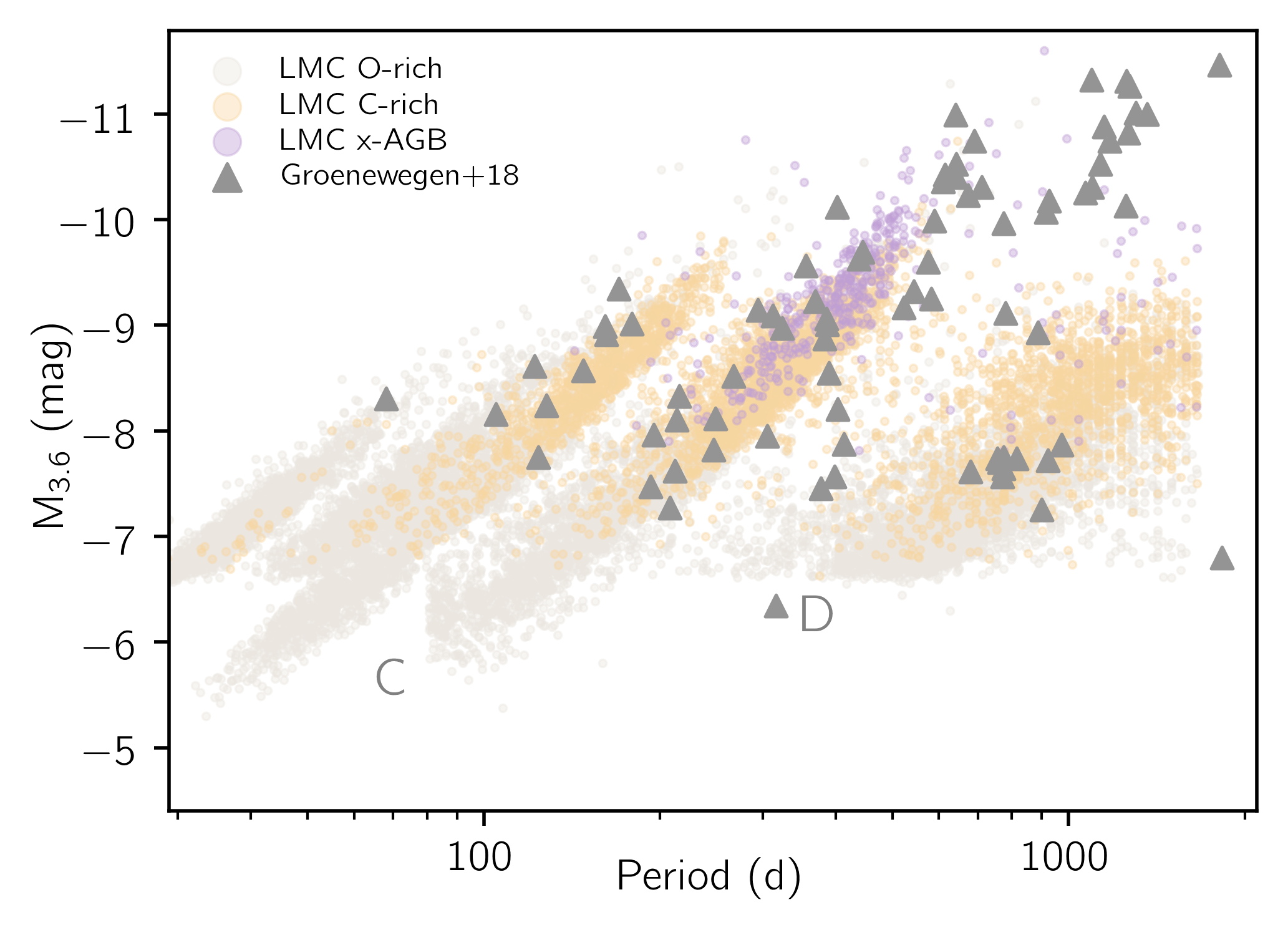}
    \caption{Period-Luminosity diagrams for oxygen-rich AGB stars in our GB sample (left) as well as our comparison sample in the Magellanic Clouds from \citet[][right]{Groenewegen2018}. Both panels show synthetic IRAC 3.6\mum\ photometry created using the best-fit \dusty\ models. For the GB sample we also include unshrouded models differing only by the optical depths at 10\mum\ which are set to 0.1. In the background we show the MACHO+SAGE sample from \citet{Riebel2010} containing oxygen- and carbon-rich AGB stars, as well as more evolved and dusty x-AGB stars. We have also labeled sequences C and D, associated with the fundamental mode and LSP, respectively; for the LMC, we adopt a distance modulus of $M$--$m = 18.52$ mag \citep{Kovacs2000}.}
    \label{fig:p_l_relation}
\end{figure*}

\begin{figure}
    \centering
    \includegraphics[width=\columnwidth]{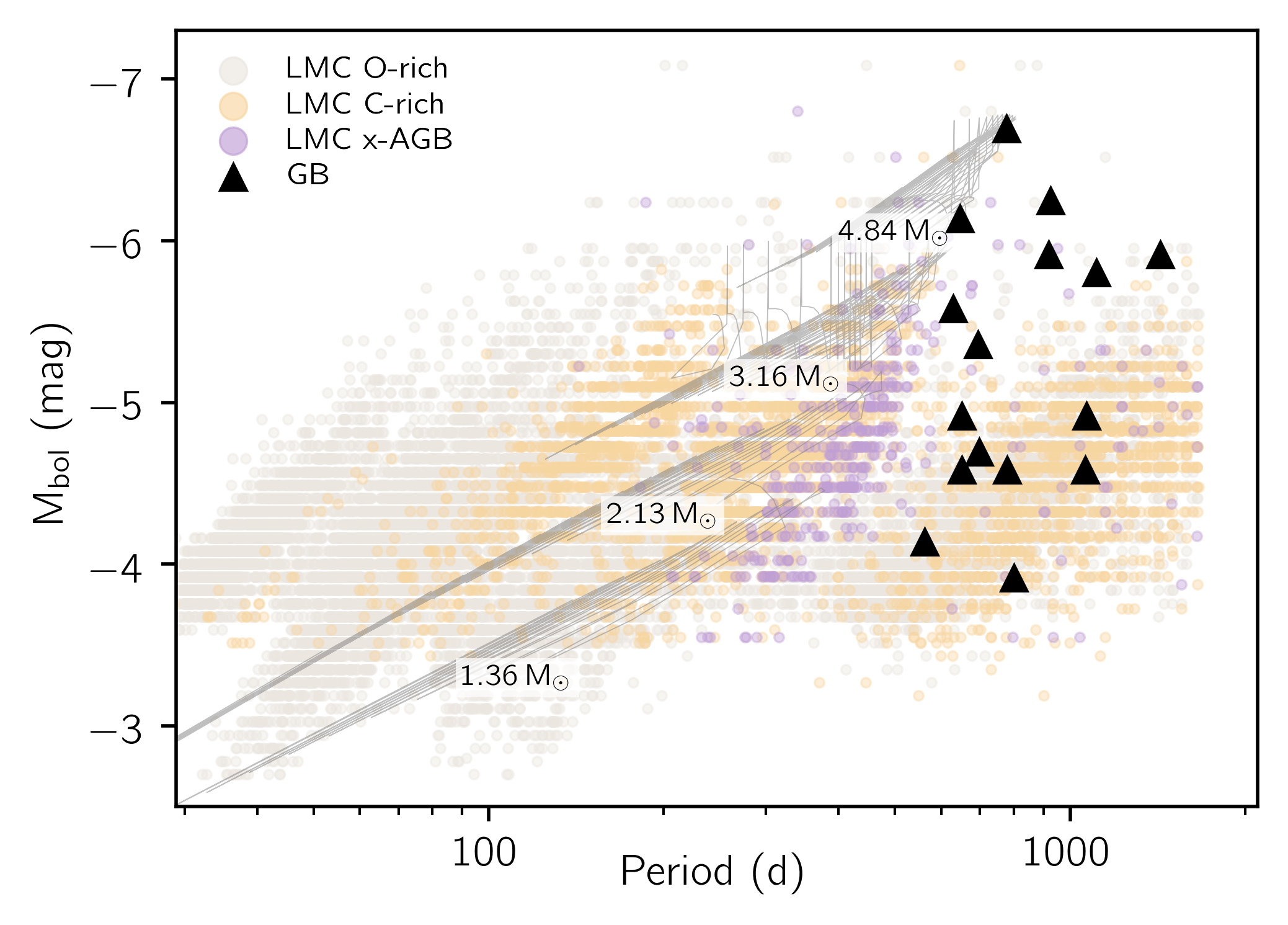}
    \caption{$P$--$L$ diagram for oxygen-rich AGB stars in our GB sample using bolometric magnitudes; symbols follow the same convention as Figure \ref{fig:p_l_relation}. Also shown are example PARSEC \citep{Bressan2012,Marigo2013} isochrones of solar-metallicity TP-AGB stars with ages ranging from 4\,--\,0.125\,Gyr and initial masses ($M_{\rm init}$) from $\sim$1--5\,M$_{\odot}$ \citep{Marigo2008}; isochrones are based on nonlinear pulsation models from \citet{Trabucchi2021}.}
    \label{fig:mbol_p_l_relation}
\end{figure}

\subsection{Pulsation Properties} \label{sec:pulsation_properties}

Our sample was selected on the basis of IR color and variability. We now have pulsation periods for 16 of our 21 sources. These were determined using near-IR data from the VVV survey \citep{Goldman2017,Blommaert2018,Molnar2022}, WISE \citep{Groenewegen2022}, and ESO's 2.2 and 3.6 m telescopes \citep{vanderVeen1990}. Previous surveys \citep[e.g., the Optical Gravitational Lensing Experiment; ][]{Soszynski2013} may have missed these sources as they are faint at optical wavelengths.

Mira variables are primarily fundamental-mode pulsators, falling mainly on sequence C of the $P$--$L$ diagram \citep{Wood1999}. While using mid-IR magnitudes can lessen the effects of dust, mid-IR $P$--$L$ relations still show a broadening of these sequences as a result of circumstellar dust \citep{Goldman2019a} as well as the mass loss and evolution of the star \citep{Trabucchi2017}; this evolution includes a complicated interplay of changes in surface gravity, effective temperature, current mass, and chemistry (M. Marengo, private communication). When plotted on an IR $P$--$L$ diagram, most of the available pulsation periods for the GB sample fall on sequence D associated with the LSP (Figure \ref{fig:p_l_relation}).

As our sources are far dustier than the sample in the Magellanic Clouds, to understand the effects of increased line-of-sight circumstellar dust on the interpretation of the pulsation behavior of our sample, we have modeled ``unshrouded'' photospheric SEDs for our GB sample. Using the same model properties as our best-fit \dusty\ models except with $\tau_{10} = 0.1$, we retrieve models that represent the stars without significant circumstellar dust, similar to the dust levels in the Magellanic Clouds sample.\footnote{We attempt to reproduce dust levels ($\tau_{0.5\mu{\rm m}} \sim 1.0$) seen in the sample by \citet{Groenewegen2018}. These are best reproduced by our \dusty\ models using $\tau_{10\mu{\rm m}} \sim 0.1$.} We then produce synthetic 3.6\mum\ IRAC photometry using these and the best-fit models by integrating them along the throughput of the detector. The resulting unshrouded best-fit synthetic photometry is shown in Figure \ref{fig:p_l_relation}.

While we see a spread in the best-fit synthetic photometry between sequences C and D, several sources are clearly seen to migrate away from sequence D in the unshrouded photometry. The two brightest unshrouded sources (IRAS 17367$-$3633 and IRAS 17316$-$3523) are likely fundamental-mode pulsators, where the second brightest is appearing as an LSP star in the IR. We see several more migrating toward sequence C as we remove the effects of the enhanced circumstellar dust.

Figure \ref{fig:mbol_p_l_relation} shows the $P$--$L$ relation instead using bolometric magnitudes ($M_{\rm bol}$) to better visualize the targets' integrated flux, at the cost of less well-defined sequences. The two brightest unshrouded sources (IRAS 17367$-$3633 and IRAS 17316$-$3523) are seen also on the fundamental mode when looking at bolometric magnitudes, with the rest of the sources scattered toward sequence D. The distribution of the GB sources is similar to the distribution of the unshrouded sample in Figure \ref{fig:p_l_relation}. This highlights the importance of using bolometric magnitudes for interpreting pulsation behavior. Using luminosities from optical and near-IR filters, we expect even more fundamental-mode pulsators on the LSP as the effects of circumstellar dust are more pronounced in these regimes. It is also critical to take into account luminosities when determining the pulsation mode as long periods may not necessarily imply pulsation on sequence D. Miras in the Local Group have been found to continue the fundamental-mode sequence to long periods and high luminosities \citep{Karambelkar2019}.

There are several caveats to interpreting the $P$--$L$ relation of our sources. First, the distance of our sources is unclear. The depth of the bulge is $\sim 1.4$\,kpc, resulting in an uncertainty in the absolute 3.6\mum\ magnitudes of around 1 mag. Additionally, the models are fitted to a single epoch of observations and are not phase-averaged values. While this gives an accurate snapshot of the flux of each source, these stars typically vary by at least 1 mag in the mid-IR, adding to the uncertainty in the brightness.

\subsection{Indications of Equatorial Enhancement}
\label{sec:equatorial_enhancement}
Our sample shows a variety of unique characteristics that individually can be explained with several scenarios; taken as a whole, however, it suggest stars with equatorially enhanced geometries.
 
\paragraph{Silicate Dust}
The formation scenario for crystalline silicates typically involves reaching sufficient temperatures to anneal the dust grains \citep[1067~K;][]{Hallenbeck2000}, without reaching a temperature where they are vaporized (2000~K). That said, evidence has shown crystalline formation at lower temperatures \citep{Molster1999}. We suspect that these silicates are either formed in a clumpy or equatorially enhanced outflow. 

Theoretical evidence predicts that higher gas densities promote the formation of crystalline silicate dust \citep{Tielens1998,Gail1999,Sogawa1999}. Clumpy outflows with densities higher than a typical outflow have been observed in high-resolution studies of nearby AGB stars \citep{Ohnaka2016,Stewart2016,Wittkowski2017}, and modeled using 3D radiation-hydrodynamics codes \citep{Hoefner2019}. Alternatively, several works have demonstrated that crystalline silicates form in equatorially enhanced circumstellar outflows \citep{Molster1999,Edgar2008}. We suspect that our sample follows this scenario as it exhibits the IR characteristics of two known equatorially enhanced OH/IR stars, OH\,26.5+0.6 and OH\,30.1--0.7. These sources, which are likely to be undergoing or recently underwent HBB, show similar crystalline and amorphous silicate absorption features as our sample \citep{Marini2023}\footnote{While \citet{Decin2020} show evidence of equatorial enhancement, \citet{Marini2023} find that they do not need to invoke binarity to explain the observed IR characteristics of the system.}. Additional observations of CO with the Atacama
Large Millimeter/submillimeter Array have detected shell-like spiral structures in these two sources likely induced by a wide-binary companion \citep{Mastrodemos1999,Decin2019}.

\paragraph{Expansion Velocities}
Our modeled \dusty\ expansion velocities are far lower than the values measured using OH masers. While several interpretations are discussed in \S \ref{sec:pulsation_properties}, the discrepancy may indicate that we are probing different parts of the circumstellar environment. As noted by \citet{Decin2019}, we measure the warm dust in the IR and the cool gas in the radio. We may be seeing the difference between an asymmetric inner dust region \citep[shaped by a companion; see][]{Decin2019}, and a less perturbed outer wind that is more uniform on larger scales \citep[see ][]{Ramstedt2020}.

\paragraph{Pulsation Properties}
Our sample may not be true LSP stars, or current models of the LSP may not be sufficient to explain our sample. \citet{Trabucchi2017} demonstrate that when the dominant pulsation period is the LSP, theoretical models show the LSP stars tend to have secondary periods in the first radial overtone mode (sequences B and C\arcmin). Between these sequences ($\sim$ 60 days), the onset of significant mass loss begins \citep{McDonald2019}, as well as the LSP beginning to dominate as the primary mode. This may be indicating a point at which a specific range of orbital radii can sustain the dusty subsolar close-binary companions proposed by \citet{Soszynski2021}. At larger orbital radii, a star may not be able to accrete significant material, while at smaller orbital radii, the companion is absorbed. Additionally, there may be AGB luminosities that restrict the LSP. This is consistent with the observation that the LSP amplitude is smaller for stars with low and high luminosity as opposed to moderate luminosity \citep{Percy2023}. 

Most of the sources that we observe on sequence D may be a result of mass loss and an extended atmosphere, and indirectly a result of a binary companion. \citet{Vassiliadis1993} outline a scenario in which stars evolve roughly horizontally from the fundamental mode, on the $M_{\rm bol}$ $P$--$L$ relation. High mass-loss rates would result in a reduction of the envelope density and consequently longer periods at roughly the same luminosity \citep{Vassiliadis1993}. Additionally, a wide-binary companion would allow for an even lower envelope density over the surface but concentrated along an equatorial plane. Models have shown that pulsation periods observed in extended circumstellar envelopes can be longer than the pulsation period of the stellar photosphere \citep{Bowen1988}. We expect that at least two sources in our sample appear as LSP stars in the mid-IR as a result of dust and a filter bias. For the remainder, we suspect that significant mass loss and an extended equatorially enhanced circumstellar outflow shaped by a companion (e.g. a wide binary) can explain the LSP. In this scenario, the stars exhibit fundamental-mode pulsations longer than equally luminous Miras on sequence C. Long periods of $\sim$1590 \citep{Engels2015proceedings,Suh2002,vanlangevelde1990} and $\sim$2000 days \citep{Olivier2001,Groenewegen2022,vanlangevelde1990} are also observed in the previously mentioned equatorially enhanced OH/IR stars OH\,26.5+0.6 and OH\,30.1--0.7, respectively. 
 
The fraction of these more obscured LSP sources is likely lower than the LSP sources proposed by \citet{Soszynski2021}. The latter may also be more similar to the Magellanic Cloud LSP sources that are more centrally located on sequence D. Sources like those in our sample reach higher luminosities and may have been more preferentially studied in previous IR variability surveys. They also span a large range of luminosities and masses. As a result, these dusty LSP sources may have contributed to the confusion surrounding the origin of the LSP. 

\section{Conclusion}
We have targeted a sample of highly obscured OH/IR stars in the GB using low-resolution mid-IR spectroscopy. The sample was selected for being variable and bright in the IR, and we consistently see strong absorption in the 10\mum\ silicate and 11.3\mum\ crystalline silicate features. We have modeled the SEDs of the sample and estimated luminosities, wind speeds, and mass-loss rates that are similar to previous results. The discrepancy between the observed and modeled wind speeds may suggest high drift velocities, high gas-to-dust ratios, or point to the regions of different densities being probed in the inner (IR) and outer (radio/submillimeter) circumstellar envelope. While similar to previous results, we suspect that the measured mass-loss rates are inflated for this sample as a result of nonspherically symmetric circumstellar envelopes. The 16 sources with measured pulsation periods
appear associated with the enigmatic LSP on the IR $P$--$L$ diagram. By modeling the sources without significant dust, at least two sources appear closer to sequence C associated with fundamental-mode pulsators. The remaining sources likely migrated from the fundamental mode to longer periods with the same luminosities as they underwent intense mass loss. An equatorial enhancement in the circumstellar geometry would also result in longer periods as the envelope density decreases except for a higher-density band along an equatorial plane. The high dust content, crystalline silicate absorption, discrepancy in the inner and outer measured wind speeds, periods longer than fundamental-mode pulsators of equal luminosity, and SEDs and pulsation periods similar to those of known equatorially enhanced OH/IR stars suggest to us that these stars are also equatorially enhanced. Simulations of the observed characteristics of similar OH/IR stars have been explained using circumstellar envelopes that are shaped by wide-binary companions; this is also a plausible explanation for our heavily obscured AGB sample. 

\section*{Acknowledgments}
We thank Rachel Street for combing through the ROME/REA catalog for us. Based on observations collected at the European Organisation for Astronomical Research in the Southern Hemisphere under ESO program 099.D-0907(A).
O.C.J. acknowledges support from an STFC Webb fellowship. 

\vspace{5mm}
\facility{VLT:Melipal \citep[VISIR;][]{Lagage2004}}
\software{\desk\ \citep{Goldman2020}, \dusty\ \citep{Elitzur2001}, astropy \citep{astropy:2013,astropy:2018}, numpy \citep{numpy2011,numpy2020}, scipy \citep{scipy2020}, pyphot \citep{pyphot}, dust\_extinction \citep{Gordon2024}}

\bibliographystyle{aasjournal}
\bibliography{references_2024}

\appendix
\section{Additional SED figures}
Here we include additional zoomed-in versions of the best-fit models, photometry, IRS spectra, and new VISIR spectra; the colors used are the same as in Figure \ref{fig:GB_seds}. \\
\begin{figure*}[b]
  \begin{minipage}[c]{\textwidth}
  \begin{center}
     \includegraphics[height=3.9cm, trim={0.2cm 0.8cm 0.2cm 0}, clip]{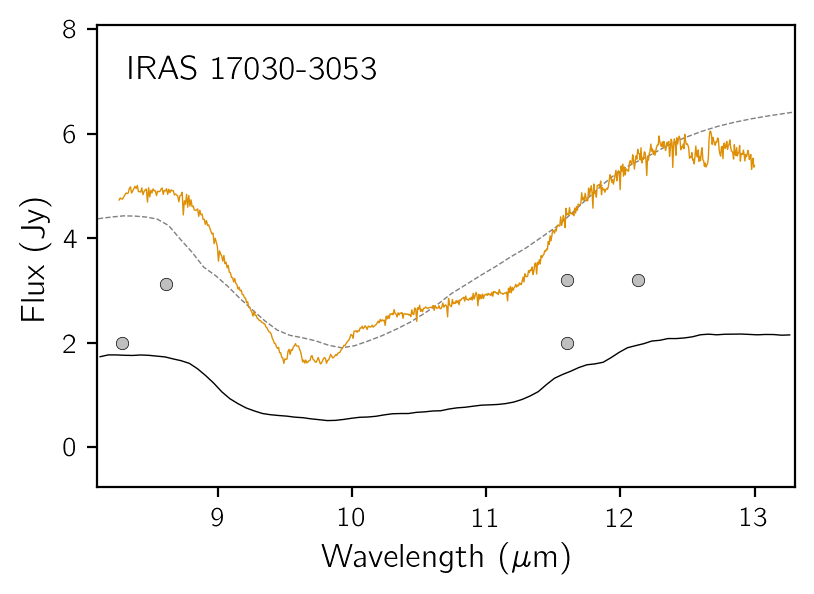}
     \includegraphics[height=3.9cm, trim={0.7cm 0.8cm 0.2cm 0}, clip]{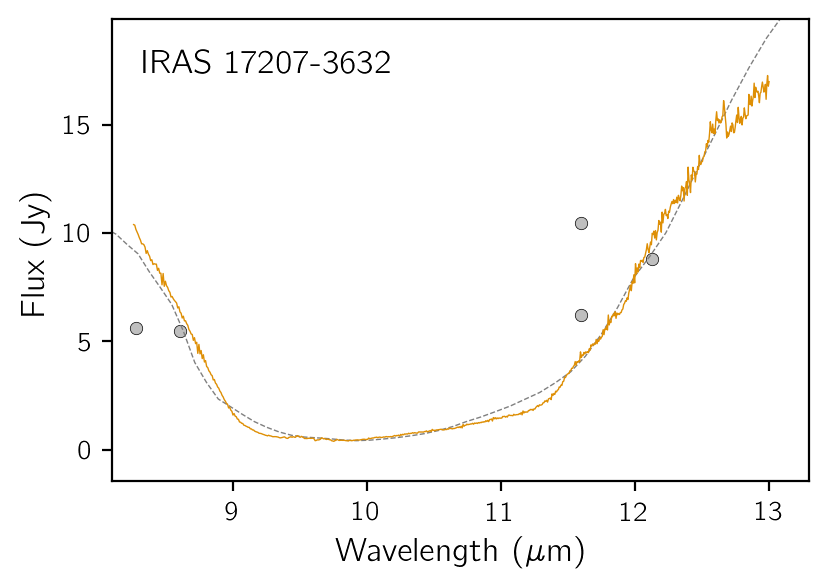}
     \includegraphics[height=3.9cm, trim={0.7cm 0.8cm 0.2cm 0}, clip]{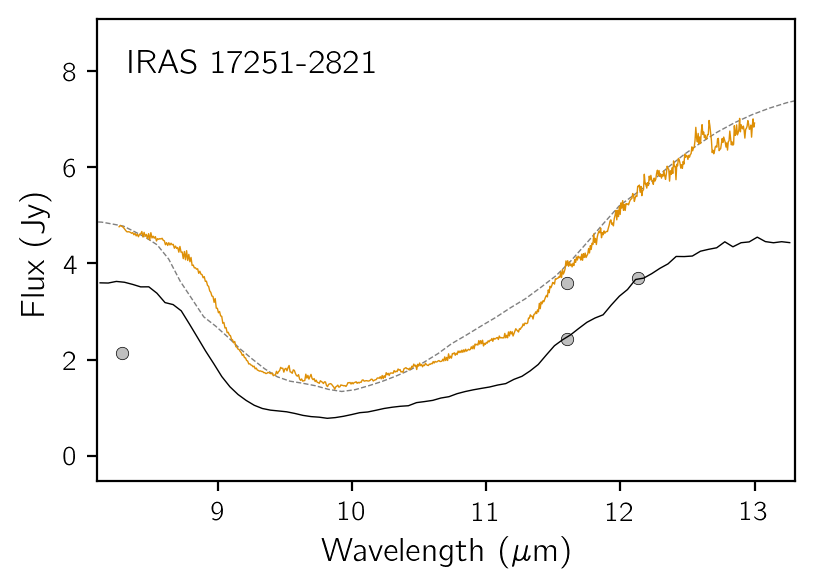}
     \includegraphics[height=3.9cm, trim={0.2cm 0.8cm 0.2cm 0}, clip]{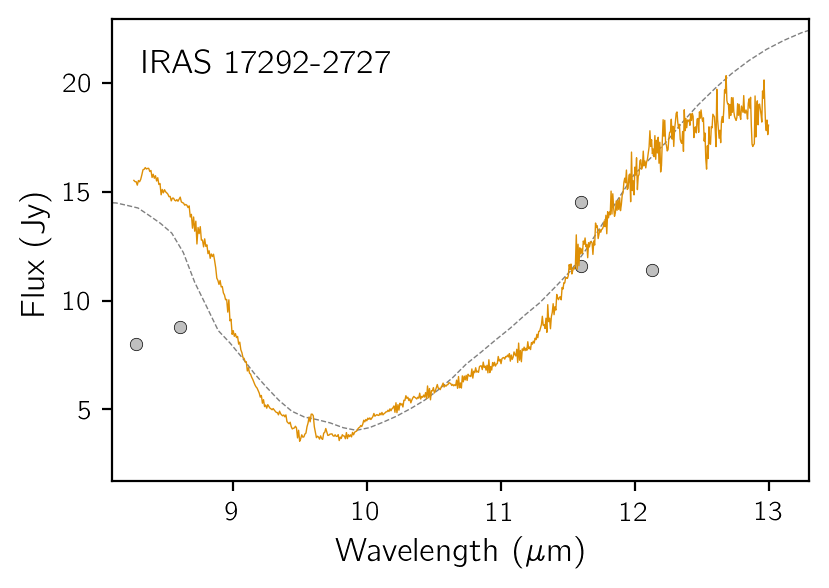}
     \includegraphics[height=3.9cm, trim={0.7cm 0.8cm 0.2cm 0}, clip]{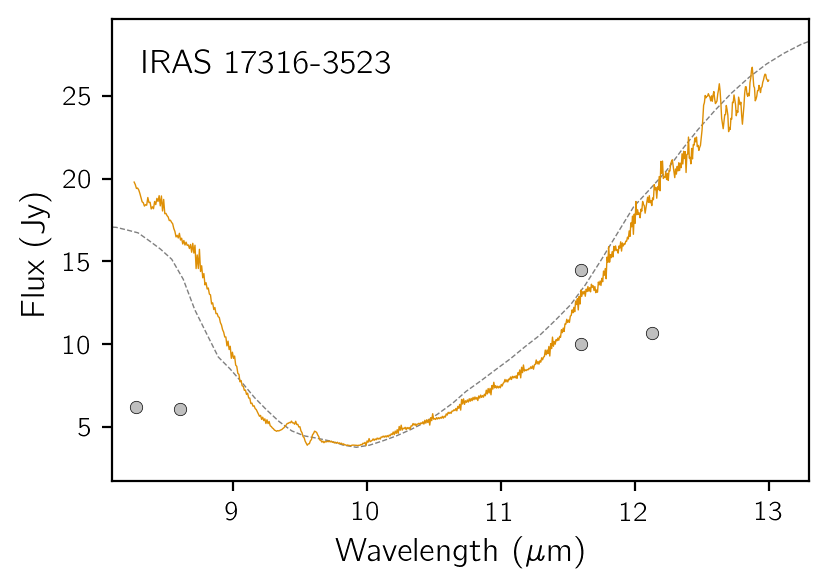}
     \includegraphics[height=3.9cm, trim={0.7cm 0.8cm 0.2cm 0}, clip]{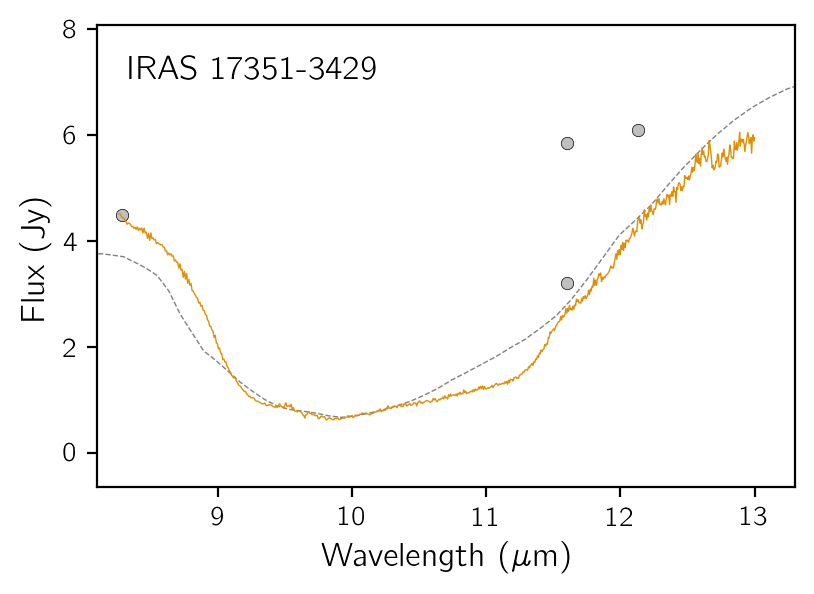}
     \includegraphics[height=3.9cm, trim={0.2cm 0.8cm 0.2cm 0}, clip]{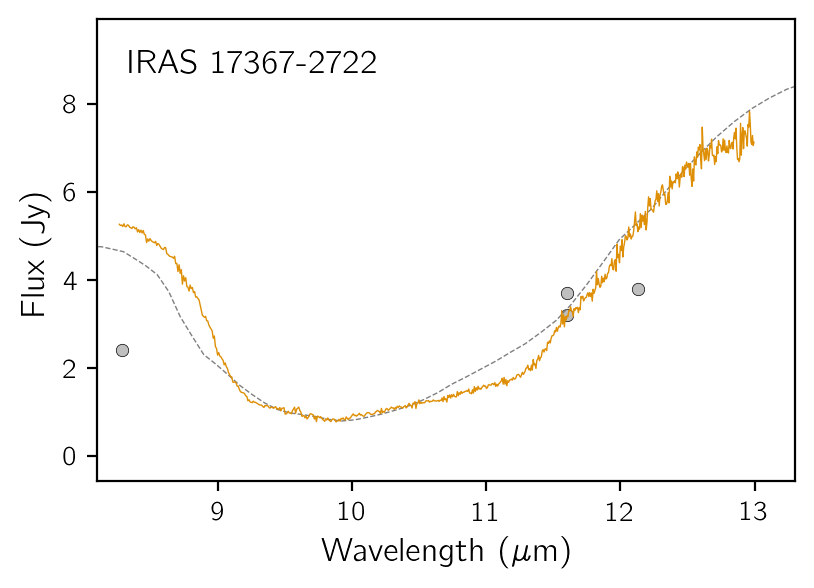}
     \includegraphics[height=3.9cm, trim={0.7cm 0.8cm 0.2cm 0}, clip]{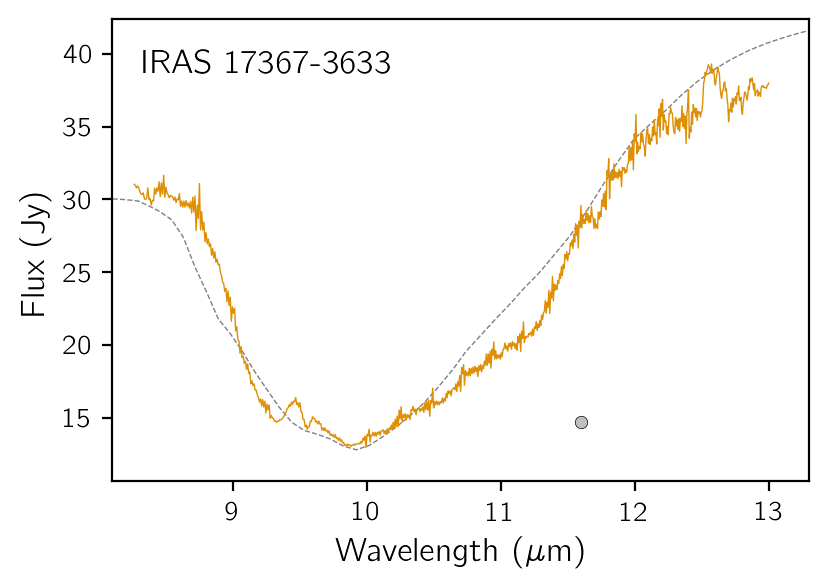}
     \includegraphics[height=3.9cm, trim={0.7cm 0.8cm 0.2cm 0}, clip]{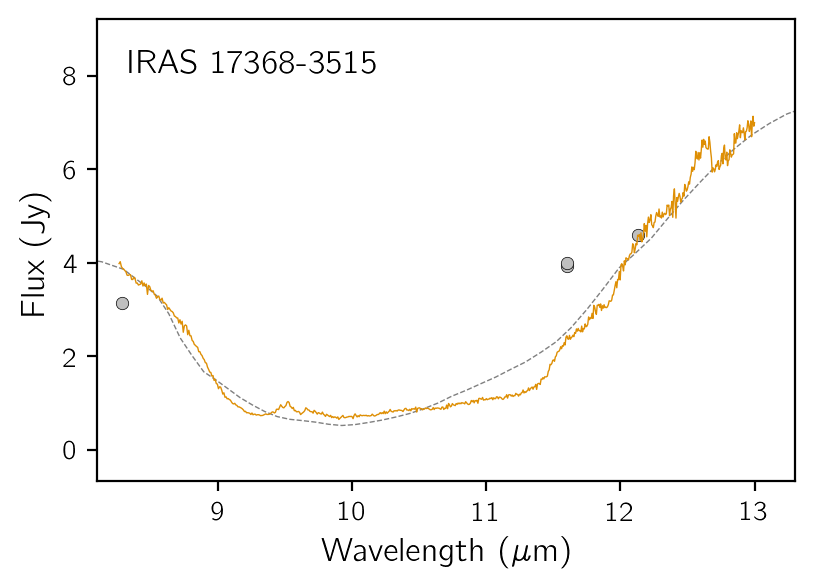}
     \includegraphics[height=4.375cm, trim={0.2cm 0.0cm 0.2cm 0}, clip]{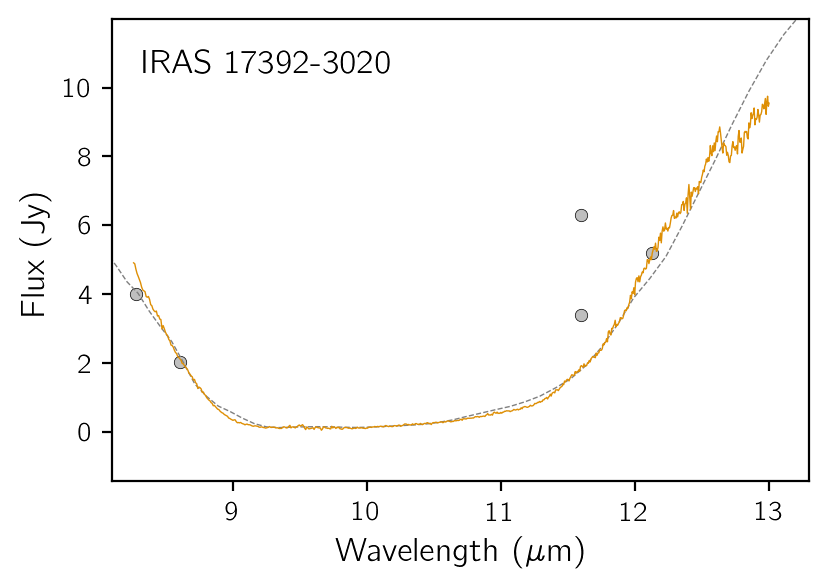}
     \includegraphics[height=4.375cm, trim={0.7cm 0.0cm 0.2cm 0}, clip]{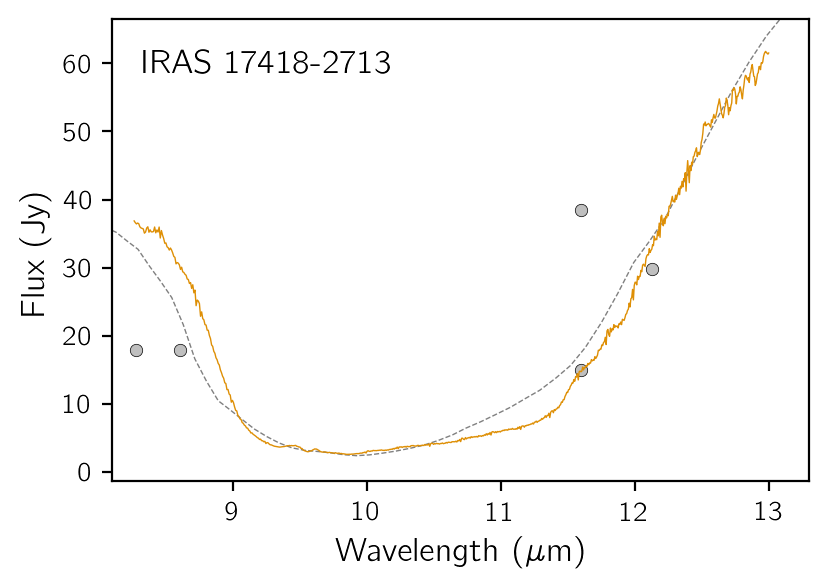}
     \includegraphics[height=4.375cm, trim={0.7cm 0.0cm 0.2cm 0}, clip]{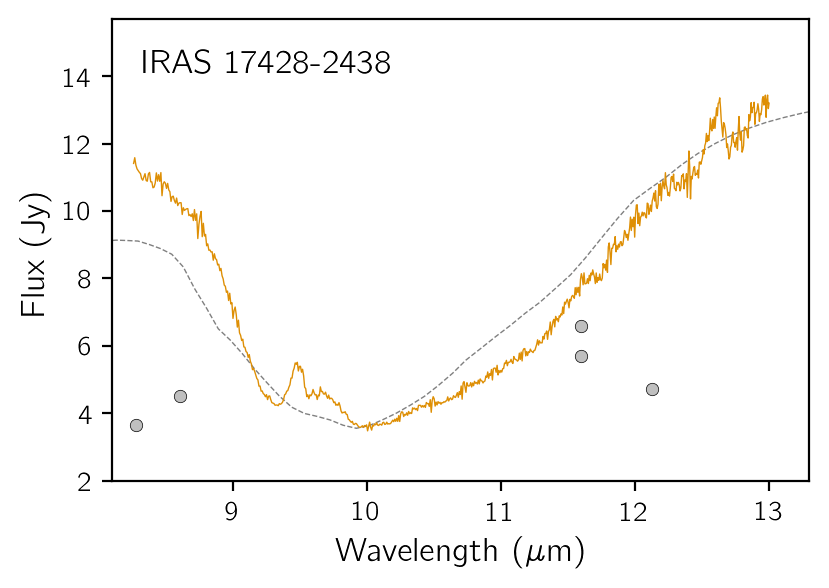}
   \end{center}
   \caption{The SED fitting of \dusty\ models (dashed line) to VISIR ( orange) spectra for our sources within the GB. Also shown are the available \spitz\ IRS spectra (in thin black) and available mid-IR photometry from \citet{Jimenez-Esteban2015}.}
   \label{fig:GB_seds_zoom_10}
   \end{minipage}
\end{figure*}

\setcounter{figure}{7}
\begin{figure*}
  \begin{minipage}[c]{\textwidth}
  \begin{center}
     \includegraphics[height=3.9cm, trim={0.2cm 0.8cm 0.2cm 0}, clip]{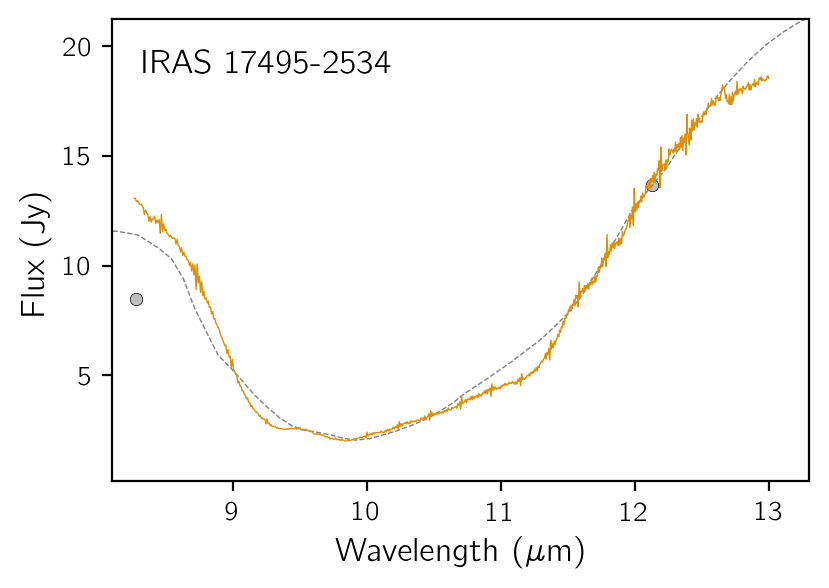}
     \includegraphics[height=3.9cm, trim={0.7cm 0.8cm 0.2cm 0}, clip]{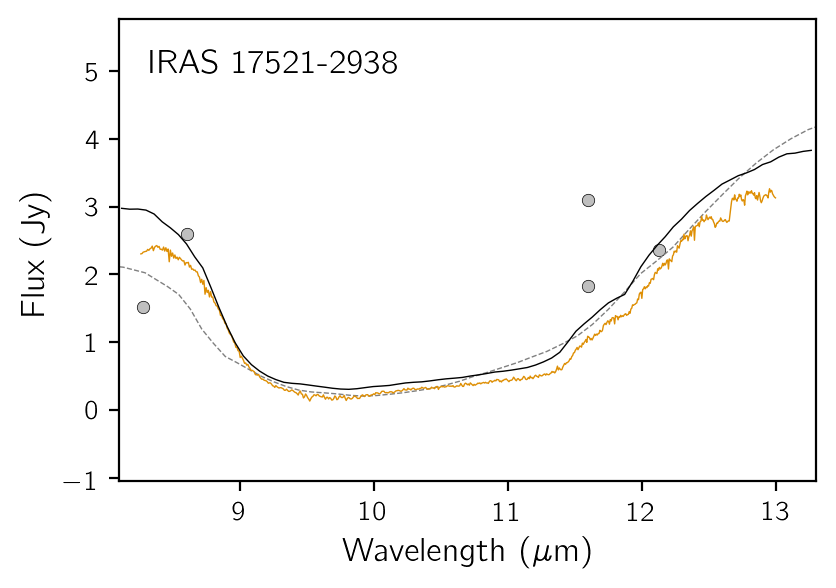}
     \includegraphics[height=3.9cm, trim={0.7cm 0.8cm 0.2cm 0}, clip]{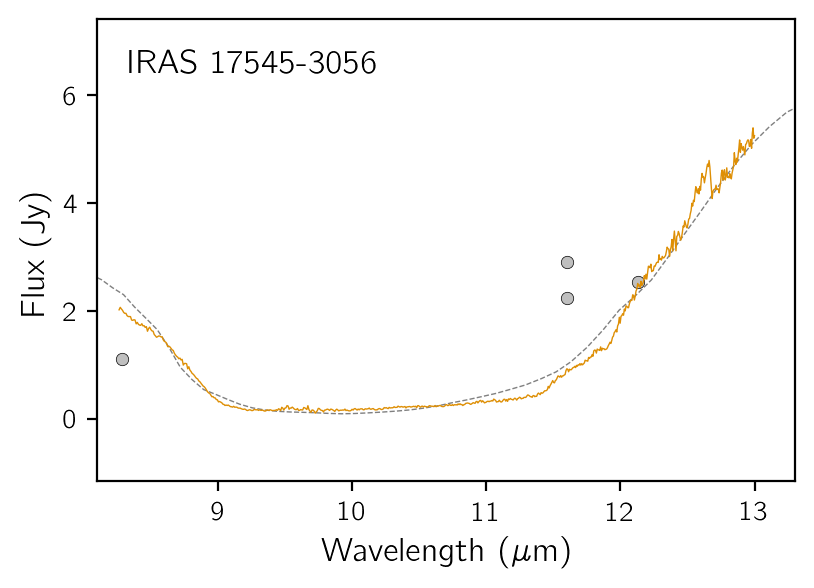}
     \includegraphics[height=3.9cm, trim={0.2cm 0.8cm 0.2cm 0}, clip]{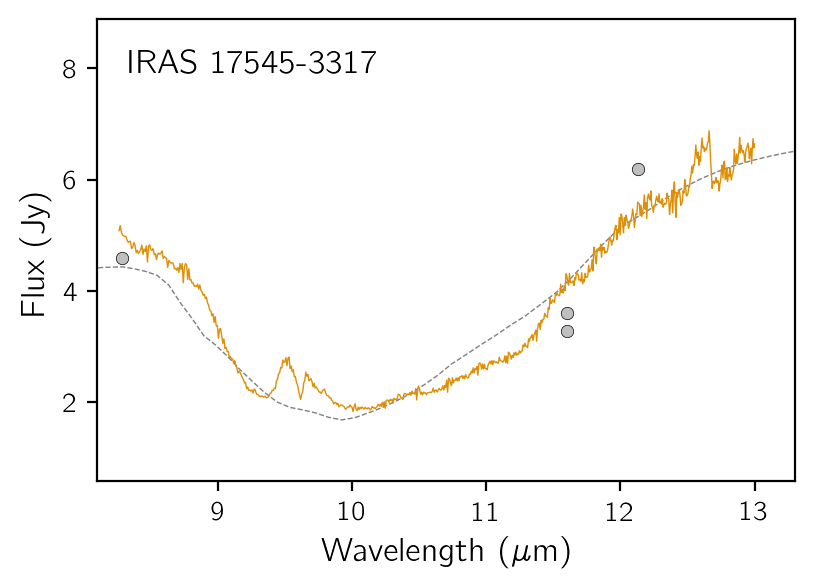}
     \includegraphics[height=3.9cm, trim={0.7cm 0.8cm 0.2cm 0}, clip]{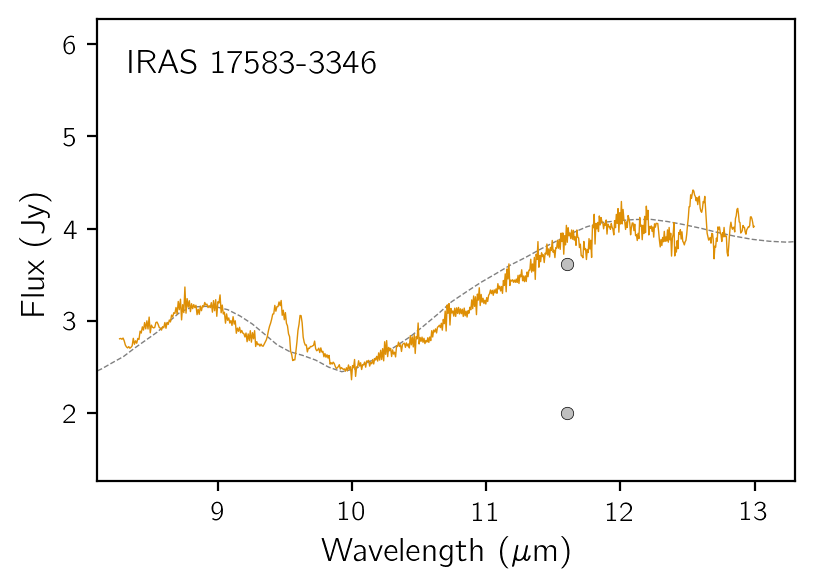}
     \includegraphics[height=3.9cm, trim={0.7cm 0.8cm 0.2cm 0}, clip]{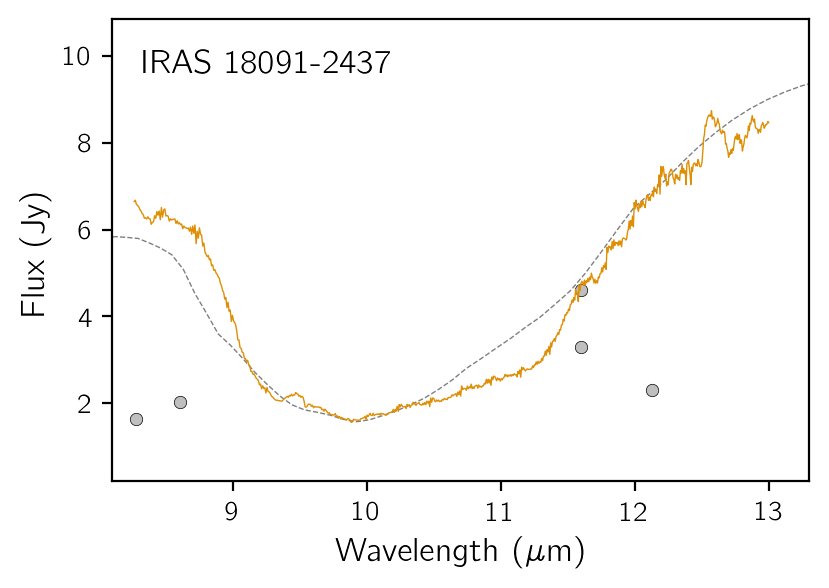}
     \includegraphics[height=4.375cm, trim={0.2cm 0.0cm 0.2cm 0}, clip]{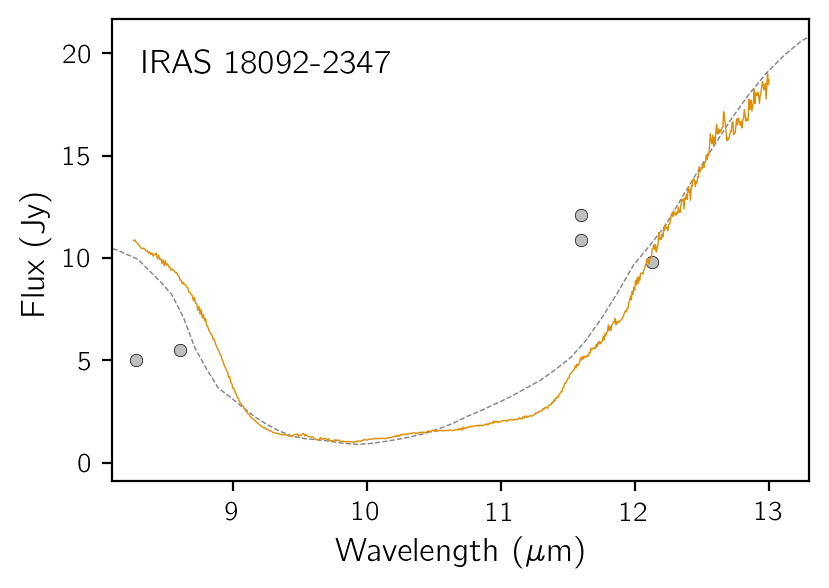}
     \includegraphics[height=4.375cm, trim={0.7cm 0.0cm 0.2cm 0}, clip]{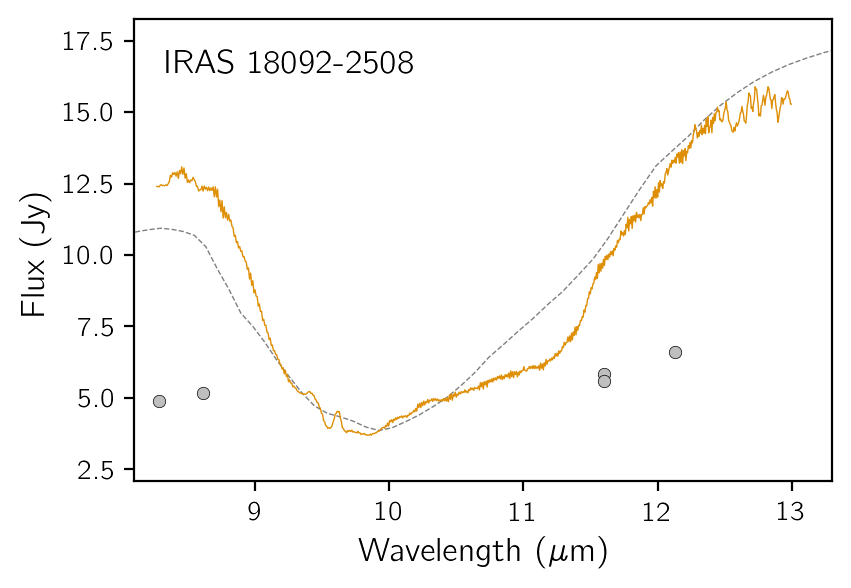}
     \includegraphics[height=4.375cm, trim={0.7cm 0.0cm 0.2cm 0}, clip]{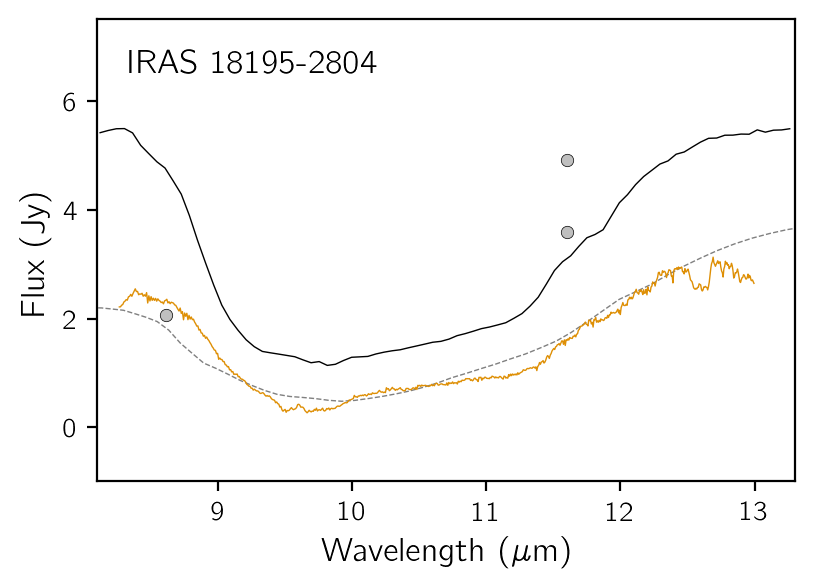}
   \end{center}
   \caption{continued}
   \end{minipage}
\end{figure*}
\label{lastpage}
\end{document}